\documentclass[aps,prl,reprint,superscriptaddress,longbibliography]{revtex4-2}
\pdfoutput=1

\usepackage[T1]{fontenc}
\usepackage[utf8]{inputenc}
\usepackage{amsmath,amssymb,amsthm,mathtools,graphicx,booktabs}
\usepackage{mathrsfs}
\usepackage[expansion=false]{microtype}
\usepackage[colorlinks=true,citecolor=blue,linkcolor=blue,urlcolor=blue]{hyperref}
\usepackage[capitalise,noabbrev]{cleveref}
\usepackage{orcidlink}

\newcommand{\dd}{\text{d}}
\newcommand{\Veff}{V_\text{eff}}

\newcommand{\Sext}{S_\text{ext}}

\hypersetup{pdfauthor={Kevin Goldstein, Vishnu Jejjala, Suresh Nampuri}}
\hypersetup{pdftitle={Entropy Products and the Extremal Limit That Never Comes}}

\begin{document}

\title{Entropy Products and the Extremal Limit That Never Comes}

\author{Kevin Goldstein\,\orcidlink{0000-0002-7872-0580}}
\email{kevin.goldstein@wits.ac.za}
\affiliation{Mandelstam Institute for Theoretical Physics, School of Physics, University of the Witwatersrand, Johannesburg, South Africa}
\affiliation{National Institute for Theoretical and Computational Sciences (NITheCS), Gauteng, South Africa}

\author{Vishnu Jejjala\,\orcidlink{0000-0003-2603-6717}}
\email{vishnu.jejjala@gmail.com}
\affiliation{Mandelstam Institute for Theoretical Physics, School of Physics, University of the Witwatersrand, Johannesburg, South Africa}
\affiliation{National Institute for Theoretical and Computational Sciences (NITheCS), Gauteng, South Africa}

\author{Suresh Nampuri\,\orcidlink{0000-0003-1378-4644}}
\email{nampuri@gmail.com}
\affiliation{Center for Mathematical Studies, Faculdade de Ci\^encias, Universidade de Lisboa, Edf.\ C6, Campo Grande, 1749-016 Lisboa, Portugal}

\begin{abstract}
Two regular black hole horizons need not have an entropy product fixed by the charges and couplings.
In Einstein--Maxwell--scalar theory, we construct asymptotically flat branches with products above or below the squared extremal entropy.
Regularity selects their horizon scalar values, and different branches can share the charges, couplings, and asymptotic scalar.
Each product stays constant as mass and asymptotic scalar vary.
The horizon areas stay apart and no extremal limit is reached: the temperature vanishes only where the mass diverges.
We prove local existence.
\end{abstract}

\maketitle

Extremal black holes exhibit the attractor mechanism~\cite{Ferrara:1995ih,Strominger:1996kf,Sen:2005wa,Goldstein:2005hq}.
At fixed charges, the horizon scalars settle at a critical point of an effective potential, making the entropy independent of the asymptotic scalars.
Cveti\v{c}, Gibbons, and Pope found that products of horizon areas depend only on quantized charges and angular momenta in broad families of black holes~\cite{Cvetic:2010mn}.
The product of the two physical horizon areas depends on mass in Reissner--Nordstr\"om--AdS~\cite{Visser:2012wu}, as does the product of Wald entropies in some higher curvature theories~\cite{Castro:2013pqa}.
In the asymptotically flat theories considered here, a family with two regular horizons and a smooth limit to the extremal attractor obeys~\cite{Larsen:1997ge,Cvetic:1997uw,Cvetic:1997xv,Cvetic:2010mn,Goldstein:2014gta}
\begin{equation}
 S_+S_-=\Sext^2\,.
 \label{eq:usual-product}
\end{equation}
This relation can be thought of as the equivalent of the attractor mechanism at finite temperature (hot attractor)~\cite{Goldstein:2014gta,Goldstein:2015pcl,Goldstein:2018mwt}, with $\Sext$ denoting the extremal entropy at the same charges and couplings~\footnote{
The definition in Ref.~\cite{Goldstein:2018mwt} uses the zero temperature limit of the family; here, $\Sext$ is set by the theory and charges.
The definitions agree when that family has a smooth extremal limit at the same attractor.}.
The individual entropies vary with mass and asymptotic scalars.
In this Letter, we examine whether the characterization of the product in terms of charges alone extends across disconnected regular branches and show that it does not.

Constancy and normalization are distinct statements.
The two horizon first laws and radial equations make the product constant on a connected regular family at fixed charges and couplings.
A smooth connection to the extremal solution supplies the additional step that fixes this constant to $\Sext^2$~\cite{Goldstein:2014gta,Goldstein:2015pcl}.
Davydov found regular dilaton branches over coupling windows with discrete horizon scalar values~\cite{Davydov:2017zbs}.
These are selected trajectories among generic singular interiors, rather than the only black holes with flowing scalars.
He distinguished them from the known Toda families, in which the horizon scalar varies continuously, and argued that the fixed coupling extremal limit of the discrete branches is singular.

\begin{figure*}[t]
 \centering
 \includegraphics[width=\textwidth]{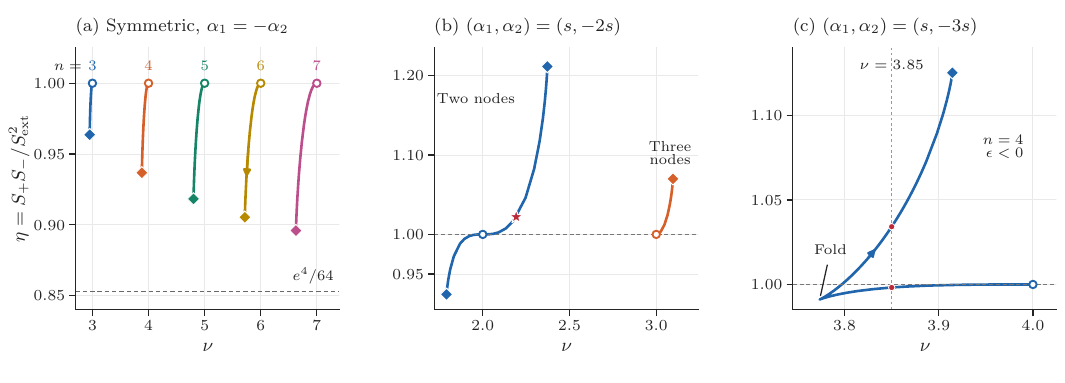}
\caption{\textsf{Entropy products on regular branches, with $\epsilon=\phi_+-\phi_0$ the deviation of the outer horizon scalar from its attractor value.
(a) Symmetric couplings $\alpha_1=-\alpha_2$ with $Q_1=Q_2=1$, labeled by the scalar node count $n$.
The samples at $\epsilon=6$ reach the endpoint diamonds to the displayed precision.
(b) Couplings $(\alpha_1,\alpha_2)=(s,-2s)$ with $(Q_1^2,Q_2^2)=(2,1)$, showing two and three nodes.
The star marks the fixed theory $(\sqrt7,-2\sqrt7)$.
(c) The $\epsilon<0$ side of the four node branch for couplings $(\alpha_1,\alpha_2)=(s,-3s)$ and $(Q_1^2,Q_2^2)=(3,1)$, with the fold marked.
The dotted line at $\nu=3.85$ intersects a deficit orbit below unity and an excess orbit above it.
Open circles denote Reissner--Nordstr\"om limits, diamonds denote endpoint predictions from separated reflections, and arrows indicate increasing $|\epsilon|$.
Each point represents an orbit.
Note that for all values of $\nu$, there is an $\eta=1$ constant scalar solution.}}
 \label{fig:products}
\end{figure*}

We show that regularity permits constant entropy products above or below that of the extremal solution.
Four-dimensional Einstein--Maxwell--scalar theory with positive scalar kinetic energy and no scalar potential admits continuous asymptotically flat families with two regular horizons for which
\begin{equation}
 S_+S_-=C=\eta\Sext^2\,,\qquad \eta>0\,,\quad\eta\ne1\,.
 \label{eq:product-shift}
\end{equation}
Along each such family, which we call an orbit, the mass and asymptotic scalar vary together while the product remains fixed; varying the coupling traces a branch of orbits.
Fixing the charges, couplings, and asymptotic scalar selects at most one member of each orbit.
Distinct orbits can nevertheless reach the same charges, couplings, and asymptotic scalar with different products and masses.
The product therefore requires a branch label even after the asymptotic scalar is specified.
The ordinary first laws and Smarr relations hold on these stationary families.
We establish deficits ($\eta<1$) over open intervals of dilaton theories with equal and opposite couplings and extend the construction to couplings of unequal magnitude, where excesses ($\eta>1$) also occur.
Following the branches numerically, we find deficits above $10\%$ and excesses above $20\%$ (Figure~\ref{fig:products}).
A positive coupling response explains the signs and constrains folds with different products at the same coupling and node count.
We also show that none of these orbits reaches an extremal limit at finite mass, and we describe the finite area end of each orbit, where both horizon areas stay finite.
Supplemental Material~\cite{Supplement} gives additional derivations, numerical checks, and extensions.

The shift in the product reflects a competition between scalar focusing and displacement from the attractor.
Scalar gradients lower the product of the horizon radii relative to the outer horizon potential, while displacement from the minimum raises that potential above its attractor value.
In a family connected to the extremal solution these effects cancel in the product.
On the branches studied here, inner horizon regularity selects isolated values of the horizon scalar away from the attractor value.
Generic values instead give a Kasner singularity~\cite{Davydov:2017zbs}.
This singular case connects to the interior mechanics of Xiong and L\"u~\cite{Xiong:2026zkb}.
Regularity selects both the horizon scalar and the product normalization.
Constancy here concerns stationary variations preserving both regular horizons, not evolution at fixed asymptotic scalar.
Regular inner horizons may be destabilized by mass inflation~\cite{Poisson:1990eh}, and dynamical stability remains to be established.

\noindent\textit{First laws and entropy products.}\quad
Consider a canonical neutral scalar and two electric fields with action
\begin{equation}
 I=\frac1{16\pi}\int\!\dd^4x\,\sqrt{-g}
 \left[R-2(\nabla\phi)^2-\sum_{a=1}^2e^{-\alpha_a\phi}(F^a)^2\right]\,,
 \label{eq:action}
\end{equation}
where $G=1$, the dilaton couplings satisfy $\alpha_1>0>\alpha_2$, and both charges are nonzero.
Eliminating the Maxwell fields gives
\begin{equation}
 \Veff=Q_1^2e^{\alpha_1\phi}+Q_2^2e^{\alpha_2\phi}\,,
 \qquad \Sext=\pi\Veff^0\,,
 \label{eq:potential}
\end{equation}
where $\phi_0$, the unique minimum of $\Veff$, is the attractor value of the horizon scalar, $e^{(\alpha_1-\alpha_2)\phi_0}=-\alpha_2Q_2^2/(\alpha_1Q_1^2)$, and $\Veff^0=\Veff(\phi_0)$.
We refer to $\alpha_1=-\alpha_2$ as the symmetric model.
The constant scalar Reissner--Nordstr\"om family, with $\phi\equiv\phi_0$ and $r_+r_-=\Veff^0$, exists at every coupling and has $\eta=1$.
For solutions with two nondegenerate horizons and metric $\dd s^2=-a^2\dd t^2+\dd r^2/a^2+b^2\dd\Omega_2^2$, the radial equations and Hamiltonian constraint respectively imply~\cite{Goldstein:2005hq}
\begin{equation}
 \begin{gathered}
 \mathcal R:=a^2b^2=(r-r_+)(r-r_-)\,,\quad \Delta=\tfrac{1}{2}(r_+-r_-)>0\,,\\
 b''=-b(\phi')^2\,,\qquad
 (\mathcal R\phi')'=\frac{\partial_\phi\Veff}{2b^2}\,.
 \end{gathered}
 \label{eq:radial}
\end{equation}
and
\begin{equation}
	1-\frac{\mathcal R'b'}b+\frac{\mathcal R(b')^2}{b^2}
	=\frac{\Veff}{b^2}-\mathcal R(\phi')^2\,.
	\label{eq:Hradial}
\end{equation}
Primes denote radial derivatives.
We write $\phi_\pm=\phi(r_\pm)$ and $b_\pm=b(r_\pm)$ for the horizon values of the scalar and the radius, and $\epsilon=\phi_+-\phi_0$ for the deviation of the outer scalar from its attractor value.
We use the asymptotic normalization $a^2\to1$, $b/r\to1$ when defining the mass and thermodynamic potentials.

For variations through stationary solutions with both horizons regular, the closed Iyer--Wald surface form gives a first law at each horizon~\cite{Wald:1993nt,Iyer:1994ys}:
\begin{equation}
 \delta M+\Sigma\,\delta\phi_\infty
 =T_h\delta S_h+\sum_a\Phi_h^a\delta Q_a\,,
 \qquad h=\pm\,.
 \label{eq:full-firstlaw}
\end{equation}
Here, $M$ is the asymptotic mass, $\phi=\phi_\infty+\Sigma/r+\cdots$, $S_h=\pi b_h^2$, and $T_h=(a^2)'(r_h)/(4\pi)$, with $T_-<0$.
The potentials $\Phi_h^a=-A_t^a(r_h)$ use $A_t^a(\infty)=0$; the scalar work term is common to both horizons~\cite{Gibbons:1996af}.
At fixed charges and couplings, subtracting the first laws~\eqref{eq:full-firstlaw} gives the first relation in~\eqref{eq:firstlaw}, while the radial equations~\eqref{eq:radial} give the second,
\begin{equation}
 T_+\delta S_+=T_-\delta S_-\,,
 \qquad 2T_\pm S_\pm=\pm {\Delta}\,,
 \label{eq:firstlaw}
\end{equation}
and hence $\delta\log(S_+S_-)=0$~\cite{Goldstein:2014gta}, so that $S_+S_-=C$ for some constant $C$ on each connected regular family.
Since $\Sext$ depends only on the charges and couplings, $\eta=C/\Sext^2$ is constant too.

Length scaling gives the usual Smarr relation $M=2T_hS_h+\sum_a\Phi_h^aQ_a$ at both horizons, with no scalar term~\cite{Gibbons:1996af,Castro:2012av}.
Scaling and a constant scalar shift fix the charge dependence of $C$; its dimensionless normalization $\eta$ depends only on the couplings and branch.

At fixed charges and couplings, different values of $C$ distinguish disconnected regular families.
For $\eta\ne1$, no path preserving both regular horizons reaches the extremal attractor at the same charges and couplings.
At the finite area end of each orbit, temperature vanishes while mass and asymptotic scalar diverge and the horizon entropies remain unequal.

\noindent\textit{The physics of $\eta$.}\quad
For a nonconstant scalar profile, define the positive focusing integral between the horizons
\begin{equation}
	\mathcal F_\phi
	=b_+\int_{r_-}^{r_+}\!\dd r\,(r-r_-)(\phi')^2b>0\,.
    \label{eq:Fdef}
\end{equation}
Using the equation of motion for $b$, \eqref{eq:radial}, together with the Hamiltonian constraint~\eqref{eq:Hradial} at $r_+$, we find
\begin{equation}
 \label{eq:focusing}
	C=S_+S_-=\pi^2b_+^2b_-^2=\pi^2\left(V_{\text{eff}}(\phi_+) - \mathcal{F}_\phi\right)^2\,,
\end{equation}
assuming the inner horizon is regular.
The product is determined by the balance between $\Veff(\phi_+)$ and $\mathcal F_\phi$, whose relative sign follows from~\eqref{eq:Hradial}.

\noindent\textit{A sign rule.}\quad
On a nonconstant regular branch, write the couplings as $\alpha_a=s\bar\alpha_a$, with $s>0$ and fixed $\bar\alpha_a$.
At fixed charges and $\bar\alpha_a$, varying $s$ in the first law gives
\begin{equation}
 \frac{\dd\log\eta}{\dd s}
 =\frac2{s\Delta}\int_{r_-}^{r_+}\!\dd r\,|\mathcal R|(\phi')^2>0\,.
 \label{eq:coupling-sign}
\end{equation}
Thus $\eta$ increases with $s$ on each regular branch segment parametrized by that coupling.
The index $\nu$, defined by $\nu(\nu+1)=-\alpha_1\alpha_2/2$, also increases with $s$.
A monotone segment leaving the constant scalar Reissner--Nordstr\"om family toward smaller $\nu$ has a deficit; one leaving toward larger $\nu$ has an excess.

\noindent\textit{A deficit at fixed coupling.}\quad
In the symmetric model $\alpha_1=-\alpha_2=\alpha$ with unit charges, $\Veff=2\cosh(\alpha\phi)$, so the attractor value is $\phi_0=0$ and $\Sext=2\pi$.
We expand in $\epsilon$ about the constant scalar Reissner--Nordstr\"om solution, $\phi=\phi_0+\epsilon\phi_1+\epsilon^3\phi_3+\cdots$; the even orders vanish because $\Veff$ is even about $\phi_0$.
Regularity at the outer horizon gives the Legendre function $\phi_1=P_\nu(z)$, where $z$, defined in the End Matter, maps the horizons to $\pm1$~\cite{Goldstein:2014gta}.
This solution has a logarithmic singularity at the inner horizon unless $\nu$ is an integer.
Continuing the expansion to third order, we find that the coefficient of the logarithm at the inner horizon is $\epsilon\sin(\pi\nu)/\pi+c_3\epsilon^3$, where $c_3(\nu)$ collects the cubic scalar self-interaction and the quadratic metric response~\cite{Supplement}.
If $\sin(\pi\nu)$ is of order $\epsilon^2$, we can tune $\epsilon$ to cancel the singularity and obtain a regular inner horizon.
For small $\epsilon$ this is possible only for $\nu$ close to an integer, and on one side of it.
Since $c_3=0$ at $\nu=1$ and $2$, the logarithm in $\phi_1$ cannot be cancelled in this way near these values.
The first branch therefore leaves Reissner--Nordstr\"om at $\nu=3$, where $c_3=-96/1001$.
Writing $\delta_\nu=3-\nu$, so that $\sin(\pi\nu)/\pi\simeq\delta_\nu$, the leading terms are
\begin{equation}
 \epsilon^2=\frac{1001}{96}\delta_\nu+O(\delta_\nu^2)\,,
 \qquad
 1-\eta=\frac{1001}{96}\delta_\nu^2+O(\delta_\nu^3)\,,
 \label{eq:fixed-deficit}
\end{equation}
and, since $c_3<0$, regular solutions exist only below $\nu=3$.
Equivalently, $1-\eta=(96/1001)\epsilon^4+O(\epsilon^6)$.

The metric and electromagnetic responses cancel at order $\epsilon^2$, while their order $\epsilon^4$ remainder gives
\begin{equation}
 \frac{\mathcal F_\phi-[\Veff(\phi_+)-\Veff^0]}{\Veff^0}
 =\frac{48}{1001}\epsilon^4+O(\epsilon^6)>0\,.
 \label{eq:quartic-balance}
\end{equation}
This is why focusing wins on this branch.
The End Matter proves that the cancellation is exact: for some $\delta_0>0$, every fixed $\nu\in(3-\delta_0,3)$ admits a unique small $\epsilon>0$ with an exactly regular inner horizon, $\phi_-=-\phi_+$, and $\eta<1$.
It also derives the coefficients and the asymptotically flat completion.

We follow numerically the branches that leave Reissner--Nordstr\"om at $\nu=n=3,\dots,7$; each keeps the $n$ scalar nodes of its linear mode $P_n$, as in~\cite{Davydov:2017zbs}.
The curves are monotone in coupling, with limiting deficits growing with $n$ to about $10\%$ [Figure~\ref{fig:products}(a)].

\noindent\textit{Asymmetric couplings can reverse the sign.}\quad
For $\alpha_1\neq-\alpha_2$, $\Veff$ is no longer even about $\phi_0$, and the expansion in $\epsilon$ acquires even orders, $\phi=\phi_0+\epsilon\phi_1+\epsilon^2\phi_2+\cdots$, with $\phi_1=P_\nu(z)$ as before.
The coefficient of the logarithm at the inner horizon is now $\epsilon\sin(\pi\nu)/\pi+c_2\epsilon^2+\ldots$, where $c_2(\nu)$ is proportional to $\Veff'''(\phi_0)$~\cite{Supplement}.
The singularity can therefore be cancelled already at second order, with $\sin(\pi\nu)$ of order $\epsilon$.
Near a positive even integer $n$, where $c_2\neq0$, this requires
\begin{equation}
 \begin{gathered}
 \nu-n=\kappa_n\epsilon+O(\epsilon^2)\,,\\
 \kappa_n=-\frac{\Veff'''(\phi_0)}{8\Veff^0}
       \int_{-1}^1\!\dd z\,P_n(z)^3\ne0\,,
 \end{gathered}
 \label{eq:asym-crossing}
\end{equation}
with the couplings in $\kappa_n$ evaluated at $\nu=n$.
Since the tuned $\epsilon$ is linear in $\nu-n$, regular solutions now exist on both sides of the integer.
The first such branches leave Reissner--Nordstr\"om at $\nu=2$, where the symmetric model has none [Figure~\ref{fig:products}(b)].
The sign identity~\eqref{eq:coupling-sign} gives $\dd\log\eta/\dd\nu=2\epsilon^2+O(\epsilon^3)$ near $\nu=n$, and hence
\begin{equation}
 \eta=1+\frac{2}{3\kappa_n^2}(\nu-n)^3+O((\nu-n)^4)\,.
 \label{eq:asym-eta}
\end{equation}
Thus every fixed ratio $\alpha_2/\alpha_1\neq-1$ has local intervals with an excess above an even integer and a deficit below it.
The sign depends on $\nu-n$, independently of the sign of $\kappa_n$.
The End Matter shows that this cancellation is also exact, with $\phi_-=\phi_+$.
These local intervals are analytic results, as are the symmetric deficit intervals up to $n=20$.
The numerical curves in Figure~\ref{fig:products} extend these local results to finite $\epsilon$ and approach the coupling endpoints calculated below.

For example, the theory $(\alpha_1,\alpha_2)=(\sqrt7,-2\sqrt7)$ with $(Q_1^2,Q_2^2)=(2,1)$ has $\phi_0=0$ and $\Sext=3\pi$, and regularity selects
\begin{equation}
 \phi_+=\phi_-\simeq0.442794\,,\qquad
 \eta\simeq1.02205\,,
 \label{eq:excess}
\end{equation}
the star in Figure~\ref{fig:products}(b); here, the electromagnetic increase exceeds focusing.
Three independent numerical methods agree on $\eta$ to $2\times10^{-12}$~\cite{Supplement}.

\noindent\textit{Branch endpoints and folds.}\quad
Between the horizons, the variables $\dd r/\dd\tau=\mathcal R$ and $U=\frac12\log(-a^2)$ turn the radial equations into the motion of a particle in the $(U,\phi)$ plane with potential $\sum_aQ_a^2e^{2U+\alpha_a\phi}$.
It enters from the outer horizon with velocity $(\dot U,\dot\phi)=(\Delta,0)$, where dots denote $\tau$ derivatives, and is turned back by the two exponential walls.
Regularity at the inner horizon requires an exit antiparallel to the entry velocity, with $\dot\phi\to0$ and $\dot U\to-\Delta$.
At large $|\epsilon|$ the motion is well approximated by free flight between separated reflections from the walls~\cite{Supplement}.
For a symmetric branch with $n$ scalar nodes, let $\alpha_\infty$ and $\nu_\infty$ denote the limiting coupling and index as $|\epsilon|\to\infty$, and set $\beta=\pi/(2n+2)$; we call $(\nu_\infty,\eta_\infty)$ the branch endpoint.
In this approximation, the $n+1$ reflection angles fix
\begin{equation}
 \alpha_\infty=2\cot\beta\,,\qquad
 \nu_\infty(\nu_\infty+1)=\frac{\alpha_\infty^2}{2}\,.
 \label{eq:endpoint-coupling}
\end{equation}
Each exact single exponential reflection also displaces the asymptotic trajectory.
These displacements fix the relative offset of the incoming and outgoing metric asymptotes, and hence the horizon product:
\begin{equation}
 \eta_\infty(n)=\frac14\exp\Bigl[-4\sin\beta
 \sum\nolimits_{j=1}^{n+1}\hat p_j\log\hat p_j\Bigr]\,,
 \label{eq:endpoint-product}
\end{equation}
where $\hat p_j=\sin[(2j-1)\beta]$ is the incoming normal velocity in units of $\Delta$.
Its large $n$ limit is $e^4/64\simeq0.853096$, a limiting deficit of $14.7\%$ for the symmetric sequence.
The coupling formula gives $\nu_\infty\simeq12.06$ and $12.96$ for $n=13$ and $14$, reproducing Davydov's numerical thresholds for closing gaps and overlapping windows~\cite{Davydov:2017zbs}.
Convergence of the smooth branches to this limit is numerically supported, with global separation estimates still unproved.
The same reflection construction gives the asymmetric endpoints marked in Figure~\ref{fig:products}(b,c)~\cite{Supplement}.
At a large $|\epsilon|$ endpoint, $S_+\to\infty$ and $S_-\to0$ at finite product; every finite $\epsilon$ solution on the traced branch has two regular horizons.

The sign rule also links endpoints to folds.
Along a smooth branch it gives $\dd\log\eta=K\,\dd\nu$, with $K>0$.
A branch leaving Reissner--Nordstr\"om toward smaller $\nu$ must turn if it connects to an endpoint with $\eta_\infty>1$.
Figure~\ref{fig:products}(c) shows such a turn at finite horizon scalar, where two regular orbits merge at the critical coupling $\nu_c$.
Both horizons remain regular.
For a simple fold, $\nu-\nu_c\propto(\epsilon-\epsilon_c)^2$, with fold values $\epsilon_c$ and $\eta_c$, while the sign identity makes both legs approach the same finite positive tangent, $\dd\eta/\dd\nu\to\eta_cK_c$.
Their product difference generically scales as $(\nu-\nu_c)^{3/2}$, whereas their horizon scalar difference scales as $(\nu-\nu_c)^{1/2}$; numerical continuation confirms these scalings here.
The legs thus meet in a cusp in the $(\nu,\eta)$ plane, reflecting a degeneracy of the regularity condition.
Both legs near the fold are deficits; at $\nu=3.85$ one orbit has a deficit, the other an excess.

The familiar Toda families connect continuously to Reissner--Nordstr\"om and have $\eta=1$~\cite{Goldstein:2005hq,Lu:2026lkv,Zadora:2017qeo}.
Whether a continuously variable horizon scalar at fixed coupling requires Toda integrability remains conjectural~\cite{Davydov:2017zbs}.

\noindent\textit{Orbits and an entropy gap.}\quad
Each selected solution is the seed of an orbit within its fixed theory.
Translating $\tau$ in the particle motion above maps solutions to solutions: for each $\lambda>0$ a translation takes $b_+\to\sqrt\lambda\,b_+$ and $b_-\to b_-/\sqrt\lambda$ at fixed charges, couplings, horizon scalars, and product.
The solutions on this orbit are physically distinct: the mass $M$, asymptotic scalar $\phi_\infty$, and scalar charge $\Sigma$ vary.
For sufficiently large $\lambda$ the exterior is regular and asymptotically flat; the End Matter gives the exact map and this completion.
Along each nonconstant orbit, $\phi_\infty$ varies monotonically, so fixing it selects at most one member of that orbit and fixes its mass.
Alternatively, every member can be moved to a common vacuum value $\widetilde\phi_\infty$ by the exact symmetry $\phi\mapsto\phi-\phi_\infty+\widetilde\phi_\infty$, $Q_a\mapsto e^{\alpha_a(\phi_\infty-\widetilde\phi_\infty)/2}Q_a$, which leaves the metric and product unchanged but varies the charges and their ratio.
Thus fixing the vacuum trades asymptotic scalar variation for charge variation; it does not give a continuous mass family at fixed charges.
At fixed $\phi_\infty$, locally in $(M,Q_1,Q_2)$, each isolated branch $k$ fixes $M=M_k(Q_1,Q_2;\phi_\infty)$, whereas Reissner--Nordstr\"om fixes $\phi_0(Q_1,Q_2)=\phi_\infty$.
Both impose one condition; this is not an exhaustive classification.
Scalar focusing gives a uniform positive horizon area gap on each orbit at fixed charges.

\noindent\textit{The finite area end of an orbit.}\quad
On each nonconstant orbit, $D=S_+-S_-$ increases strictly with $\lambda$, so its smallest value is a boundary infimum $D_*>0$.
Approaching this boundary produces an exterior scalar runaway and drives $T_+S_+$ to zero, while $S_\pm$ approach finite unequal limits.
Let $\mathscr S=D-D_*\to0^+$ and let $\alpha_\text{d}$ be the coupling, with its sign, of the exponential dominating the scalar runaway.
The exterior equations give
\begin{equation}
 \begin{gathered}
 p=\frac{4}{4+\alpha_\text{d}^2}\,,\qquad
 T_+\propto \mathscr S^p\,,\qquad M\propto \mathscr S^{p-1}\,,\\
 \phi_\infty=-\frac{\alpha_\text{d}p}{2}
       \log\frac{\mathscr S}{\mathscr S_i}+O(1)\,,\qquad
 \frac{\Sigma}{M}\longrightarrow-\frac{\alpha_\text{d}}2\,,
 \end{gathered}
 \label{eq:orbit-endpoint-scaling}
\end{equation}
where $\mathscr S_i>0$ is a fixed entropy scale.
In particular, $M\propto T_+^{-\alpha_\text{d}^2/4}$: zero temperature is approached at divergent mass and asymptotic scalar at fixed charges.
If time is not rescaled to keep $a^2\to1$ at infinity, the limiting metric retains regular nondegenerate horizons and develops a nonflat power law infinity.
The failure to reach a regular extremal limit is thus a degeneration of the asymptotically flat boundary conditions.
The scalar work term accounts for the mass divergence along this orbit; this limit does not describe evaporation at fixed asymptotic scalar.
At the other end of the orbit, where $S_+\to\infty$ and $S_-\to0$ at fixed product, $M\sim(8\pi T_+)^{-1}$ and the exterior approaches Schwarzschild in mass units.
Supplemental Material details these limits and the exterior geometries approached in each, with numerical validation~\cite{Supplement}.

Extremal solutions and Reissner--Nordstr\"om still exist at the same charges and couplings, but the families constructed here never reach them while both horizons stay regular.
Two regular horizons therefore define a product invariant whose value requires the regular branch as well as the charges and theory.
Deficits and excesses occur over open coupling intervals, and folds preserve this ambiguity even at fixed asymptotic scalar and node count.
The ordinary first laws remain valid; the missing information in the product is the connected regular branch.
At the endpoint couplings, the separated reflection picture also suggests one parameter families that follow a regular solution almost to the inner horizon and then end at a Kasner singularity with exponents exponentially close to their horizon values.
This is a large $|\epsilon|$ counterpart of the Einstein--Rosen bridge collapse found near the onset of scalar hair~\cite{Hartnoll:2020fhc,VandeMoortel:2021gsp}; its entropy products and interior first laws~\cite{Xiong:2026zkb} remain open.

\noindent\textit{Data availability.}\quad
The code, data, and scripts required to reproduce the results and figures in this Letter are available at: \url{https://github.com/kevin-goldstein/entropy_products}.

\noindent\textit{Acknowledgments.}\quad
VJ is supported by the South African Research Chairs Initiative of the Department of Science, Technology, and Innovation and the National Research Foundation (grant 78554).
The authors acknowledge the use of Claude (Anthropic) and Codex and GPT (OpenAI).

\nocite{Lee:1990nz,Wald:1993nt,Iyer:1994ys,Castro:2012av,Ortin:2022uxa,Elgood:2020svt,Gibbons:1996af,Astefanesei:2006sy,Harlow:2019yfa,Shi:2020csw,Zadora:2017qeo,Davydov:2017zbs,Lu:2026lkv,Anabalon:2020pez,Jockers:2004yj,Ferrara:1997tw,Galli:2011fq,Xiong:2026zkb,Gibbons:1987ps,Ceresole:2007wx,Andrianopoli:2007gt,Cremonini:2023vwf,Poisson:1990eh,Sen:2005wa,Astefanesei:2006dd,Kunduri:2013gce,Astefanesei:2007bf}

\bibliography{ref}

\onecolumngrid
\clearpage

\begin{center}\textbf{END MATTER}\end{center}

\twocolumngrid
\setcounter{equation}{0}
\renewcommand{\theequation}{A\arabic{equation}}
\renewcommand{\theHequation}{A\arabic{equation}}

\noindent\textit{Local existence at fixed coupling.}\quad
For the symmetric potential $\Veff=2\cosh(\alpha\phi)$, fix $r_\pm$, $\Delta=\tfrac{1}{2}(r_+-r_-)>0$, and $b_+>0$.
Set $\rho=r-r_+$ and $\mathcal H=\partial_\phi\Veff/(2b^2)$.
The regular horizon equations are equivalent to
\begin{equation}
\begin{aligned}
 \phi(\rho)&=\epsilon+\int_0^\rho\!\dd u\,\phi'(u)\,,\\
 \phi'(\rho)&=\frac1{2\Delta+\rho}\int_0^1\!\dd u\,\mathcal H(b(u\rho),\phi(u\rho))\,,\\
 b(\rho)&=b_++v_+\rho-\int_0^\rho\!\dd u\,(\rho-u)b(u)\phi'(u)^2\,,
\end{aligned}
\label{em:volterra}
\end{equation}
where the constraint fixes $v_+=[b_+-\Veff(\epsilon)/b_+]/2\Delta$.
On a sufficiently small interval $|\rho|<\delta<\Delta$, solve the last equation with $b>b_+/2$ and substitute into the first two.
Changing $\phi'$ changes $\phi$ by $O(\delta)$ and $b$ by $O(\delta^2)$, so the map is a contraction on a bounded ball for small $\delta$.
The contraction on holomorphic functions gives analytic horizon series and parameter dependence, and hence a regular solution for $0\le\rho<\delta$, just outside the outer horizon.

In the interior variables $\tau$ and $U$ of the main text, $\ddot U=-e^{2U}\Veff<0$ gives $U$ a unique maximum at $\tau_c$ near Reissner--Nordstr\"om, analytic in the parameters.
Since $\Veff$ is even under $\phi\to-\phi$, $F(\epsilon,\nu)=\phi(\tau_c)/\epsilon$ is even in $\epsilon$ and analytic at $\epsilon=0$.
The linear solution is the Legendre function, giving~\cite{DLMF:Legendre}
\begin{equation}
 \begin{gathered}
 F(0,\nu)=P_\nu(0)
 =\frac{\sqrt\pi}{\Gamma((1-\nu)/2)\Gamma(1+\nu/2)}\,,\\
 \partial_\nu F(0,3)=\frac23\,.
 \end{gathered}
 \label{em:matching}
\end{equation}
The implicit function theorem gives a unique even analytic curve $\nu(\epsilon)$ with $F=0$, where $\phi(\tau_c)=\dot U(\tau_c)=0$, so the solution is invariant under $\tau-\tau_c\to\tau_c-\tau$ with $\phi\to-\phi$; this turns the outer series in $e^{2\Delta\tau}$ into a regular inner series in $e^{-2\Delta\tau}$, with positive finite radius and $\phi_-=-\epsilon$; continuity gives $b>0$ in between.

To remove the dependence on $b_+$ and $\Delta$, put $x=(r-r_-)/2\Delta$, $b_0=(2/b_+)(1-x)+b_+x$, and $z=[(b_++2/b_+)x-2/b_+]/b_0$, the coordinate of $P_\nu(z)$ in the main text.
For $\mathfrak b=b/b_0$, the exact equations are
\begin{equation}
\begin{gathered}
 \mathfrak b_{,zz}=-\mathfrak b\phi_{,z}^2\,,\quad
 \partial_z[(z^2-1)\phi_{,z}]
 =\frac{\alpha\sinh(\alpha\phi)}{2\mathfrak b^2}\,,\\
 \mathfrak b(1)=1\,,\qquad
 \mathfrak b_{,z}(1)=-\frac{\cosh(\alpha\epsilon)-1}{2}\,.
\end{gathered}
\label{em:projective}
\end{equation}
Since $\dd z/\dd\tau=-\Delta(1-z^2)$, the normalized matching condition is also independent of $b_+$ and $\Delta$; one neighborhood of $\epsilon=0$ works for every finite $b_+>0$.

Near $\nu=3$, expand $\Lambda=\nu(\nu+1)=12+\Lambda_2\epsilon^2+\cdots$, $\phi=\epsilon P_3+\epsilon^3q+\cdots$, and $\mathfrak b=1+\epsilon^2\mathfrak b_2+\epsilon^4\mathfrak b_4+\cdots$.
Writing primes for $z$ derivatives gives $\mathfrak b_2''=-(P_3')^2$, $\mathfrak b_2(1)=0$, $\mathfrak b_2'(1)=-6$, hence $\mathfrak b_2=(9-9z^2+15z^4-15z^6)/8$.
The third-order equation is $L_3q=\Lambda_2P_3+48P_3^3-24\mathfrak b_2P_3$, where $L_3=\partial_z[(z^2-1)\partial_z]-12$.
Regularity requires
\begin{equation}
 \Lambda_2
 =-\frac{\int_{-1}^{1}\!\dd z\,(48P_3^4-24\mathfrak b_2P_3^2)}
 {\int_{-1}^{1}\!\dd z\,P_3^2}
 =-\frac{96}{143}\,.
 \label{em:projection}
\end{equation}
The condition $q(1)=0$ fixes the regular polynomial solution.
The fourth order equation $\mathfrak b_4''=-\mathfrak b_2(P_3')^2-2P_3'q'$ with $\mathfrak b_4(1)=0$ and $\mathfrak b_4'(1)=-(\Lambda_2+24)/2$ gives $\mathfrak b_4(-1)=-48/1001$.
Since $\mathfrak b_2(-1)=0$ and $\eta=\mathfrak b(-1)^2$, this proves the quartic deficit.
Analytic inversion of $3-\nu=(96/1001)\epsilon^2+O(\epsilon^4)$ gives the unique $\epsilon>0$ at each fixed $\nu$ sufficiently close to and below $3$.
In the fixed coupling expansion of the main text, this is $c_3=\Lambda_2/7=-96/1001$ at $\nu=3$.

For asymmetric couplings, $\tau-\tau_c\to\tau_c-\tau$ alone is still a symmetry, so a solution with $\dot\phi(\tau_c)=0$ is even about $\tau_c$ and has an exactly regular inner horizon with $\phi_-=\phi_+$.
The matching function $-\dot\phi(\tau_c)/(\epsilon\Delta)$ has linear limit $P_\nu'(0)=\nu P_{\nu-1}(0)$, with a simple zero at each positive even $n$, and the implicit function theorem again gives an analytic matching curve.
Its expansion reproduces~\eqref{eq:asym-crossing}, and the coupling identity~\eqref{eq:coupling-sign} then gives~\eqref{eq:asym-eta}.

\noindent\textit{Asymptotically flat completion.}\quad
The particle motion, continued outside $r_+$ with $U=\frac12\log a^2$, has no explicit $\tau$ dependence, so translating $\tau$ preserves the equations, charges, couplings, and horizon scalars. Since $e^{2\Delta\tau}\propto|1-x|/x$, the shift $\tau\to\tau-\log\lambda/(2\Delta)$ is a fractional linear map fixing both horizons.
For $\lambda>0$, set
\begin{equation}
\begin{gathered}
 N_\lambda=1+(\lambda-1)x\,,\qquad
 r_\lambda=r_-+\frac{2\lambda x\Delta}{N_\lambda}\,,\\
 b_\lambda(r)=\frac{N_\lambda}{\sqrt\lambda}b(r_\lambda)\,,
 \quad \phi_\lambda(r)=\phi(r_\lambda)\,,
 \quad a_\lambda^2=\frac{\mathcal R}{b_\lambda^2}\,.
\end{gathered}
\label{em:map}
\end{equation}
Since $a^2$ and $\phi$ are carried along at fixed $\tau$, $\mathcal R=a^2b^2$ and $r_\lambda'=\mathcal R(r_\lambda)/\mathcal R(r)=h^{-2}$, with $h=N_\lambda/\sqrt\lambda$, give $b_\lambda=h\,b(r_\lambda)$.
The horizon radii become $\sqrt\lambda\,b_+$ and $b_-/\sqrt\lambda$, preserving their product.

For $\lambda>1$, the full exterior $r_+\le r<\infty$ maps into $r_+\le r_\lambda<R_\lambda$, where $R_\lambda=r_++2\Delta/(\lambda-1)$.
Sufficiently large $\lambda$ gives $R_\lambda<r_++\delta$, placing this interval where the solution is regular.
At infinity,
\begin{equation}
\begin{gathered}
 b_\lambda=c_\lambda r+O(1)\,,\qquad
 a_\lambda^2=c_\lambda^{-2}+O(r^{-1})\,,\\
 \phi_\lambda=\phi(R_\lambda)+O(r^{-1})\,,\qquad
 c_\lambda=\frac{(\lambda-1)b(R_\lambda)}{2\Delta\sqrt\lambda}\,.
\end{gathered}
\label{em:asymptotics}
\end{equation}
Smoothness at $R_\lambda$ also gives the differentiated falloffs.
Since $c_\lambda>0$, the changes $\widehat t=t/c_\lambda$ and $\widehat r=c_\lambda r$, followed by a constant radial shift, give asymptotic flatness with unit lapse and unchanged charges.
The normalized separation is $2c_\lambda \Delta$, with $\widehat T_\pm S_\pm=\pm c_\lambda \Delta/2$, and varies along the orbit.
Varying $\lambda$ over this open range therefore supplies the orbit in each selected fixed theory, with two regular horizons and the same $\eta$.

\end{document}


\title{Supplemental Material for\texorpdfstring{\\}{: }Entropy Products and the Extremal Limit That Never Comes}

\author{Kevin Goldstein\,\orcidlink{0000-0002-7872-0580}}
\email{kevin.goldstein@wits.ac.za}
\affiliation{Mandelstam Institute for Theoretical Physics, School of Physics, University of the Witwatersrand, Johannesburg, South Africa}
\affiliation{National Institute for Theoretical and Computational Sciences (NITheCS), Gauteng, South Africa}

\author{Vishnu Jejjala\,\orcidlink{0000-0003-2603-6717}}
\email{vishnu.jejjala@gmail.com}
\affiliation{Mandelstam Institute for Theoretical Physics, School of Physics, University of the Witwatersrand, Johannesburg, South Africa}
\affiliation{National Institute for Theoretical and Computational Sciences (NITheCS), Gauteng, South Africa}

\author{Suresh Nampuri\,\orcidlink{0000-0003-1378-4644}}
\email{nampuri@gmail.com}
\affiliation{Center for Mathematical Studies, Faculdade de Ci\^encias, Universidade de Lisboa, Edf.\ C6, Campo Grande, 1749-016 Lisboa, Portugal}

\maketitle

This supplement develops the covariant surface calculation, higher modes, branch endpoints, numerical checks, and exterior comparisons.
The End Matter gives the local existence proof; Section~\ref{sm:exteriors} develops the geometric gap bound and endpoint.
All calculations use $G=1$, a canonical neutral scalar, positive Maxwell kinetic terms, and no scalar potential.
We retain the notation of the Letter: $\phi_0$ is the attractor value at the minimum of $\Veff$, $\Sext=\pi\Veff(\phi_0)$, $S_+S_-=C=\eta\Sext^2$, $\Delta=\frac{1}{2}(r_+-r_-)>0$, and the symmetric model has $\alpha_1=-\alpha_2$.
The accompanying reproducibility archive at \url{https://github.com/kevin-goldstein/entropy\_products} contains code, numerical records, and validation logs.
We also show the endpoint calculation, smooth equation continuation, and figure data.

\section{Covariant surface calculation and coupling variation}
\label{sm:cps}

We use the radial gauge and $\mathcal R=a^2b^2=(r-r_+)(r-r_-)$ of the Letter.
In a frame with electric and magnetic charges, set $G_a=f_{ab}\star F^b$, $Q_a=(4\pi)^{-1}\oint G_a$, and $P^a=(4\pi)^{-1}\oint F^a$.
Fields with couplings $f_1=e^{-\alpha_1\phi}$, $f_2=e^{\alpha_2\phi}$ and respective electric and magnetic charges $Q$, $P$ give $\Veff=Q^2e^{\alpha_1\phi}+P^2e^{\alpha_2\phi}$.
Dualizing the Letter's second gauge field gives this description with $Q=Q_1$ and $P=Q_2$.
\noindent\textit{Interior equations.}\quad
Write $U=\tfrac12\log|a^2|$, $\dd r/\dd\tau=\mathcal R$, and let a dot denote a $\tau$ derivative.
Then $\dot\phi=\mathcal R\phi'$ and $\dot U=b^2(a^2)'/2$, with primes denoting $r$ derivatives.
The scalar flux equation, the Einstein equations, and the Hamiltonian constraint take the same form on either side of the outer horizon when written with the signed metric function $a^2$,
\begin{equation}
 \ddot U=a^2\Veff\,,\qquad
 \ddot\phi=\tfrac12a^2\partial_\phi\Veff\,,\qquad
 \dot U^2+\dot\phi^2-a^2\Veff=\Delta^2\,.
 \label{sm:signed-equations}
\end{equation}
The exterior has $a^2=e^{2U}$.
Between the horizons $a^2=-e^{2U}$, and the mechanical Lagrangian and its conserved energy are
\begin{equation}
 W=e^{2U}\Veff\,,\qquad
 L=\dot U^2+\dot\phi^2-W\,,\qquad
 E=\dot U^2+\dot\phi^2+W=\Delta^2\,.
 \label{sm:interior-action}
\end{equation}
In this region $\tau$ increases from $-\infty$ at $r_+$ to $+\infty$ at $r_-$.
At the two regular horizons $\dot\phi\to0$, while $\dot U\to\Delta$ at $r_+$ and $\dot U\to-\Delta$ at $r_-$.
Integrating $\ddot U=-W$ and $\ddot\phi=-\partial_\phi W/2$ therefore gives the hot attractor identities
\begin{equation}
 \int_{-\infty}^{\infty}\!\dd\tau\,W=2\Delta\,,\qquad
 \int_{-\infty}^{\infty}\!\dd\tau\,\partial_\phi W=0\,.
 \label{sm:interior-averages}
\end{equation}

\noindent\textit{Horizon terms in the coupling variation.}\quad
Choose $\tau=0$ at $(r_++r_-)/2$, so that $-\mathcal R=\Delta^2\operatorname{sech}^2(\Delta\tau)$.
Denote the intercepts of the incoming and outgoing metric asymptotes by $u_0$ and $u_1$,
\begin{equation}
 U=\Delta\tau+u_0+o(1)\quad(\tau\to-\infty)\,,\qquad
 U=-\Delta\tau+u_1+o(1)\quad(\tau\to+\infty)\,.
 \label{sm:horizon-asymptotes}
\end{equation}
Since $b^2=-\mathcal R e^{-2U}$, the product is
\begin{equation}
 b_+b_-=4\Delta^2e^{-(u_0+u_1)}\,.
 \label{sm:horizon-intercepts}
\end{equation}
A translation of $\tau$ shifts $u_0$ and $u_1$ by equal and opposite amounts, preserving the sum in~\eqref{sm:horizon-intercepts}.

Let $\mu$ be a parameter in the gauge couplings and vary it along a smooth family with fixed charges and two regular horizons.
The energy constraint gives $L-\Delta^2=-2W$, so the finite on-shell action is
\begin{equation}
 I=\lim_{T\to\infty}\int_{-T}^{T}\!\dd\tau\,(L-\Delta^2)=-4\Delta\,.
 \label{sm:subtracted-action}
\end{equation}
Vary first at finite $T$, allowing $\Delta$ to vary along the family.
The Euler--Lagrange equations leave the endpoint momenta and the explicit parameter variation,
\begin{equation}
 \delta I_T=
 [2\dot U\,\delta U+2\dot\phi\,\delta\phi]_{-T}^{T}
 -4T\Delta\,\delta\Delta
 -\delta\mu\int_{-T}^{T}\!\dd\tau\,e^{2U}\partial_\mu\Veff\,,
 \label{sm:action-boundary}
\end{equation}
where $\partial_\mu\Veff$ holds $\phi$ and the charges fixed.
The horizon scalar variations are finite, so their endpoint momenta give zero in the limit.
For a nondegenerate background, the regular horizon series have exponentially decaying remainders whose variations are $O((1+T)e^{-2\Delta T})$.
At the two cuts $\delta U(-T)=-T\delta\Delta+\delta u_0+o(1)$ and $\delta U(T)=-T\delta\Delta+\delta u_1+o(1)$.
The metric endpoint term therefore cancels the term proportional to $T$ and leaves
\begin{equation}
 -4\delta\Delta
 =-2\Delta\,\delta(u_0+u_1)
 -\delta\mu\int_{-\infty}^{\infty}\!\dd\tau\,e^{2U}\partial_\mu\Veff\,.
 \label{sm:finite-variation}
\end{equation}
Taking the logarithmic variation of $S_+S_-=\pi^2(b_+b_-)^2$ in~\eqref{sm:horizon-intercepts} cancels the remaining $\delta\Delta$.
Using $e^{2U}\dd\tau=-\dd r/b^2$ in the interior gives
\begin{equation}
 \delta\log(S_+S_-)=\frac{\delta\mu}{\Delta}
 \int_{r_-}^{r_+}\!\dd r\,\frac{\partial_\mu\Veff}{b^2}\,.
 \label{sm:coupling}
\end{equation}
For the constant scalar Reissner--Nordstr\"om solution, $b=r$ and $r_+r_-=\Veff^0$, so~\eqref{sm:coupling} gives $\delta\log(S_+S_-)=2\,\delta\log\Veff^0$, as required.

\noindent\textit{Covariant surface form.}\quad
At fixed couplings and for a fixed exact symmetry $\xi$, the Iyer--Wald surface form obeys~\cite{Lee:1990nz,Wald:1993nt,Iyer:1994ys}
\begin{equation}
 \bk_\xi=\delta\bQ_\xi-\xi\cdot\Theta\,,\qquad
\dd\bk_\xi=0\,,\qquad
 \oint_{S_2}\bk_\xi=\oint_{S_1}\bk_\xi\,.
 \label{sm:closed}
\end{equation}
The variation solves the linearized equations, and any required gauge compensator is held fixed as part of the symmetry parameter.
Stokes' theorem requires an oriented hypersurface and smooth fields, not a Cauchy surface.
It therefore applies to the timelike constant $t$ hypersurface between the horizons.

We retain the common Killing normalization $\xi=\partial_t$ and signed inner temperature $T_-<0$ used in the Letter~\cite{Castro:2012av}.
Variations in static coordinates move the horizon roots and need not be smooth at the bifurcation surfaces.
A local adjustment $X_h=-(\delta\kappa_h/\kappa_h)t\partial_t+F_h(r)\partial_r$, with $F_h(r_h)=\delta r_h$, removes this coordinate singularity.
Cutoffs in disjoint horizon neighborhoods combine these adjustments into a smooth representative agreeing with the static variation near an intermediate section.
This representative need not be stationary;~\eqref{sm:closed} requires a linearized solution, not a stationary variation.

The electric and dual gauge potentials also use separate regular horizon patches.
With potentials vanishing at infinity, set $\Phi_h^a=-A_t^a(r_h)$ and $\Psi_{h,a}=\widetilde A_{ta}(r_h)$, where $\dd\widetilde A_a=G_a$ locally.
These potential differences use separate regular gauges at the bifurcation surfaces.

Magnetic variations include the patch boundary terms, which can equivalently be expressed using the dual potentials~\cite{Ortin:2022uxa,Elgood:2020svt}.
The radial surface integral is
\begin{equation}
K(r)=\frac b2\left[(a^2)'\delta b-b'\delta(a^2)-2a^2\delta b'\right]
      -\mathcal R\phi'\delta\phi-A_t^a\delta Q_a+\widetilde A_{ta}\delta P^a\,,
 \qquad K'=0\,,
 \label{sm:K}
\end{equation}
where primes denote radial derivatives.
Its scalar term vanishes at each regular horizon, and the gravitational term contains the total radius variation $\delta b+b'_h\delta r_h$.
At infinity, $\phi=\phi_\infty+\Sigma/r+\cdots$, $a^2=1-2M/r+\cdots$, and $b=r+o(1)$ give $K=\delta M+\Sigma\delta\phi_\infty$.
At each horizon, $K=T_h\delta S_h+\Phi_h\cdot\delta Q+\Psi_h\cdot\delta P$, reproducing the first laws in the Letter.
Direct differentiation of~\eqref{sm:K}, using the Hamiltonian constraint and the radial evolution equations, independently checks the conservation law.

The scalar work term need not be integrable when $\phi_\infty$ varies freely~\cite{Gibbons:1996af,Astefanesei:2006sy}.
Fixing the asymptotic modulus gives the Hamiltonian $M$, while imposing $\Sigma=\partial_{\phi_\infty}\mathcal W_\text{bdry}$ gives $M+\mathcal W_\text{bdry}$.
The common scalar work term cancels in the horizon comparison.
This use of smooth bifurcation cuts does not require a general extension of the Harlow--Wu outer boundary formalism to null hypersurfaces~\cite{Harlow:2019yfa,Shi:2020csw}.

\noindent\textit{Smarr's relation and charge dependence.}\quad
In the electric frame of the Letter, Maxwell's equation gives $(A_t^a)'=Q_a e^{\alpha_a\phi}/b^2$, while the trace reversed $tt$ Einstein equation implies
\begin{equation}
 \frac12\bigl[b^2(a^2)'\bigr]'=\frac{\Veff}{b^2}\,,\qquad
 \mathcal K_\text{S}:=\frac12b^2(a^2)'-\sum_aQ_aA_t^a\,,\qquad
 \mathcal K_\text{S}'=0\,.
 \label{sm:smarr-charge}
\end{equation}
The conserved quantity equals $M$ at infinity and $2T_hS_h+\sum_a\Phi_h^aQ_a$ at either horizon.
Static scalar gradients affect the metric and electric potentials through the other radial equations, while the electromagnetic fields supply the source in this Einstein equation.
Equivalently, the scalar modulus has zero weight under a length rescaling, accounting for its absence from the Smarr relation.

\noindent\textit{Varying the gauge couplings.}\quad
The same response follows from the covariant identity when $\mu$ varies.
For the fixed symmetry $\xi=\partial_t$ the surface form now satisfies
\begin{equation}
 \dd\bk_\xi=-\delta\mu\,\xi\cdot\partial_\mu\bL\,,\qquad
 \left.\partial_\mu\bL\right|_{\text{fields}}
 =-\frac{\partial_\mu\Veff}{8\pi b^4}\,\boldsymbol{\varepsilon}_4\,,
 \label{sm:coupling-source}
\end{equation}
with orientation $\boldsymbol{\varepsilon}_4=b^2\sin\theta\,\dd t\wedge\dd r\wedge\dd\theta\wedge\dd\varphi$.
In evaluating the Maxwell term, first differentiate the Lagrangian at fixed field strength, then substitute the solution at fixed charges.
For a gauge kinetic function $f$, the electric and magnetic contributions to $\Veff$ are $Q^2/f$ and $fP^2$, respectively.
Integrating~\eqref{sm:coupling-source} from $r_-$ to $r_+$ at fixed charges gives
\begin{equation}
 K(r_+)-K(r_-)=T_+\delta S_+-T_-\delta S_-
 =\frac{\delta\mu}{2}\int_{r_-}^{r_+}\!\dd r\,\frac{\partial_\mu\Veff}{b^2}\,.
 \label{sm:coupling-horizons}
\end{equation}
The signed horizon relation $T_\pm S_\pm=\pm\Delta/2$ turns the middle expression into $(\Delta/2)\delta\log(S_+S_-)$, recovering~\eqref{sm:coupling}.

For the common rescaling $\alpha_a=s\bar\alpha_a$, the sign of the response follows from the interhorizon identities
\begin{equation}
 \int_{r_-}^{r_+}\!\dd r\,\frac{\partial_\phi\Veff}{b^2}=0\,,\qquad
 \int_{r_-}^{r_+}\!\dd r\,\frac{\Veff}{b^2}=2\Delta\,,\qquad
 \int_{r_-}^{r_+}\!\dd r\left[
 \frac{(\phi-\phi_0)\partial_\phi\Veff}{b^2}+2\mathcal R(\phi')^2\right]=0\,.
 \label{sm:averages}
\end{equation}
The first two identities are~\eqref{sm:interior-averages} in the radial coordinate.
The third follows by multiplying $(\mathcal R\phi')'=\partial_\phi\Veff/(2b^2)$ by $\phi-\phi_0$ and integrating by parts.
Regularity gives $\mathcal R\phi'\to0$ at both horizons.
At fixed charges and fixed $\bar\alpha_a$, $s\partial_s\Veff=\phi\partial_\phi\Veff$, and the first identity allows $\phi$ to be replaced by $\phi-\phi_0$ in the integral.
The minimum value $\Veff^0$ is independent of $s$ because the potential depends on $s$ and $\phi$ through $s\phi$.
Consequently $\delta\log(S_+S_-)=\delta\log\eta$, and the third identity gives the positive response~\eqref{sm:scaling} on each nonconstant regular branch segment parametrized by $s$.

\section{Higher scalar modes and perturbative checks}
\label{sm:quartic}

The $n=3$ coefficient and its fixed coupling inversion are derived in the End Matter.
For the higher modes, use the same projective coordinate $z\in[-1,1]$ and expansions $\phi=\epsilon P_n+O(\epsilon^3)$, $b/b_0=1+\epsilon^2\mathfrak b_2+O(\epsilon^4)$, with $\epsilon=\phi_+$.
Primes in this section denote $z$ derivatives.
At $\epsilon=0$ the normalized matching functions at $z=0$ become $P_\nu(0)$ for odd $n$ and $P_\nu'(0)=\nu P_{\nu-1}(0)$ for even $n$.
Their zeros at the corresponding integers are simple.
Reflection gives $\phi_-=(-1)^n\epsilon$, so the local argument in the End Matter extends to these modes.
Let $\mathfrak b_2''=-(P_n')^2$, $\mathfrak b_2(1)=0$, and $\mathfrak b_2'(1)=-n(n+1)/2$; the cubic projection is
\begin{equation}
 c_n=-\frac{n(n+1)}2\int_{-1}^{1}\!\dd z\,
 \left(\frac{n(n+1)}{3}P_n^4-2\mathfrak b_2P_n^2\right)\,,\qquad
 \nu_n=n+c_n\epsilon^2+O(\epsilon^4)\,,\qquad
 \eta_n=1+c_n\epsilon^4+O(\epsilon^6)\,.
 \label{sm:higher}
\end{equation}
For $c_n\ne0$, the entropy coefficient also follows by integrating $\dd\log\eta/\dd\nu=2\epsilon^2+O(\epsilon^4)$ along the symmetric branch, using~\eqref{sm:scaling} and~\eqref{sm:gradient-leading}.
Exact rational evaluation gives $c_1=c_2=0$ and $c_n<0$ for $3\le n\le20$, including $c_3=-96/1001$, $c_4=-800/2431$, and $c_5=-33600/46189$.
Each negative coefficient establishes a further local interval of theories with a deficit.
The same exact coefficients satisfy $|c_3|<|c_4|<\cdots<|c_{20}|$, so at a common sufficiently small $\epsilon$ the leading displacement $n-\nu_n=|c_n|\epsilon^2+O(\epsilon^4)$ grows with node number.
This local ordering does not determine the widths of the full coupling windows at finite $\epsilon$.
The vanishing coefficients at $n=1,2$ agree with the exact $A_1\oplus A_1$ and $A_2$ Toda families, where the coupling stays fixed and $\eta\equiv1$~\cite{Zadora:2017qeo}.
The signs of the coefficients beyond $n=20$ remain undetermined by these calculations.

\noindent\textit{Logarithm coefficient at fixed coupling.}\quad
The Letter describes the same selection at fixed $\nu$, as a cancellation of logarithms at the inner horizon.
Integrating the scalar equation of the End Matter between the horizons gives the exact identity $\lim_{z\to-1}(1+z)\partial_z\phi=(\alpha/4)\int_{-1}^{1}\!\dd z\,\sinh(\alpha\phi)/\mathfrak b^2$ with $\mathfrak b=b/b_0$, so inner regularity is the vanishing of this integral.
Expand $\phi=\epsilon\phi_1+\epsilon^3\phi_3+O(\epsilon^5)$ and $\mathfrak b=1+\epsilon^2\mathfrak b_2+O(\epsilon^4)$ with $\phi(1)=\epsilon$, so that $\phi_1=P_\nu$ and $\phi_3(1)=0$; the even orders of the scalar vanish because $\Veff$ is even.
With $\Lambda=\nu(\nu+1)$ and $L_\nu=\partial_z[(z^2-1)\partial_z]-\Lambda$, the cubic order is
\begin{equation}
 L_\nu\phi_3=S_\nu\,,\qquad
 S_\nu=\frac{\Lambda^2}{3}P_\nu^3-2\Lambda\mathfrak b_2P_\nu\,,\qquad
 \mathfrak b_2''=-(P_\nu')^2\,,\quad \mathfrak b_2(1)=0\,,\quad \mathfrak b_2'(1)=-\frac{\Lambda}{2}\,.
 \label{sm:cubic-fixed-nu}
\end{equation}
Near the inner horizon $P_\nu=[\sin(\pi\nu)/\pi]\log(1+z)+O(1)$.
Writing $w=1+z$, the operator acts there as $-2\partial_w(w\partial_w)$, while $S_\nu$ diverges only as a power of $\log w$; the source therefore adds a single logarithm to $\phi_3$ and no higher powers.
Green's identity with $P_\nu(-z)$, the solution of $L_\nu u=0$ regular at the inner horizon, gives its coefficient:
\begin{equation}
 \lim_{z\to-1}(1+z)\,\partial_z\phi
 =\epsilon\,\frac{\sin(\pi\nu)}{\pi}+c_3(\nu)\,\epsilon^3+O(\epsilon^5)\,,\qquad
 c_3(\nu)=\frac12\int_{-1}^{1}\!\dd z\,P_\nu(-z)\,S_\nu(z)\,.
 \label{sm:log-coefficient}
\end{equation}
Here, the subscript denotes the order in $\epsilon$, as in the Letter.
At an integer, $P_n(-z)=(-1)^nP_n(z)$ gives $c_3(n)=(-1)^{n+1}c_n$ in terms of~\eqref{sm:higher}, and setting~\eqref{sm:log-coefficient} to zero with $\sin(\pi\nu)/\pi\simeq(-1)^n(\nu-n)$ reproduces $\nu_n=n+c_n\epsilon^2$.
The coefficient thus vanishes at $\nu=1,2$ and equals $-96/1001$ at $\nu=3$, where the scalar self-interaction and the metric response contribute $-11568/5005$ and $144/65$, respectively.
Near $\nu=1,2$ the logarithm in $\phi_1$ is present, with coefficient $\sin(\pi\nu)/\pi\simeq(-1)^n(\nu-n)$.
The higher order coefficients vanish at these Toda couplings, where every $\epsilon$ gives a regular solution, so they are also proportional to $\nu-n$ and cannot cancel it at small $\epsilon$; no branch leaves Reissner--Nordstr\"om at $n=1,2$.
At $\nu=3$ the solution of~\eqref{sm:cubic-fixed-nu} regular at the outer horizon is
\begin{equation}
 \begin{gathered}
 \phi_3=\frac{96}{1001}\,\partial_\nu P_\nu\big|_{\nu=3}+q\,,\qquad
 \partial_\nu P_\nu\big|_{\nu=3}=P_3(z)\log\frac{1+z}{2}+\frac{(z-1)(37z^2+7z-8)}{12}\,,\\
 q=\frac{z(z^2-1)(6325z^6-3215z^4+4555z^2-897)}{572}\,,
 \end{gathered}
 \label{sm:phi3}
\end{equation}
so that $\phi_3\simeq-(96/1001)\log[(1+z)/2]-32/91$ as $z\to-1$.
On the matching curve $\nu=3-(96/1001)\epsilon^2$, the first term cancels against the expansion of $\epsilon P_\nu$, leaving $\phi=\epsilon P_3+\epsilon^3q+O(\epsilon^5)$ with the odd polynomial $q$ of the End Matter, for which $q(\pm1)=0$.

\begin{figure}[tbp]
 \centering
 \includegraphics[width=.52\textwidth]{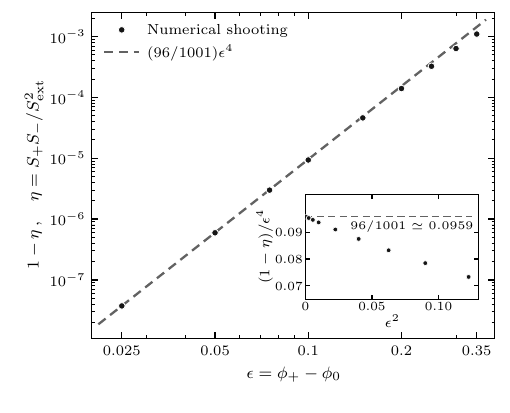}
\caption{\textsf{The symmetric branch with three scalar nodes near $\epsilon=0$.
Numerical shooting approaches the quartic coefficient $-c_3=96/1001$ in~\eqref{sm:higher}.
The corrections at finite $\epsilon$ explain why the scaled deficit at $\epsilon=0.2$ is approximately $0.087521$, rather than $0.0959041$.
Each $\epsilon$ on the curve selects a different coupling.}}
 \label{sm:fig-quartic}
\end{figure}

\section{Asymmetric couplings and the excess example}
\label{sm:asymmetric}

Take $\Veff=Q^2e^{\alpha_1\phi}+P^2e^{\alpha_2\phi}$ with $\alpha_1>0>\alpha_2$.
At its unique minimum $\phi_0$,
\begin{equation}
 \nu(\nu+1)=\frac{\Veff''(\phi_0)}{2\Veff^0}
 =-\frac{\alpha_1\alpha_2}{2}\,,\qquad
 \frac{\Veff'''(\phi_0)}{\Veff^0}
 =-\alpha_1\alpha_2(\alpha_1+\alpha_2)\,.
 \label{sm:slope-index}
\end{equation}
Set $\epsilon=\phi_+-\phi_0$.
To derive the crossing coefficient in the Letter, expand $\phi-\phi_0=\epsilon P_n+\epsilon^2q+\cdots$ and $\nu=n+\kappa_n\epsilon+\cdots$.
Since $b/b_0=1+O(\epsilon^2)$, the scalar equation at quadratic order is
\begin{equation}
 \left\{\partial_z[(z^2-1)\partial_z]-n(n+1)\right\}q
 =(2n+1)\kappa_nP_n+\frac{\Veff'''(\phi_0)}{4\Veff^0}P_n^2\,.
 \label{sm:asymmetric-source}
\end{equation}
Multiplying by $P_n$ and integrating, with $\int_{-1}^1\!\dd z\,P_n^2=2/(2n+1)$, gives the solvability condition quoted in the Letter.
The cubic integral $\int_{-1}^1\!\dd z\,P_n^3$ vanishes for odd $n$ and is positive for even $n$.
It is twice the square of the Wigner $3j$ symbol with entries $(n,n,n;0,0,0)$.
The Letter states the same condition at fixed $\nu$.
There $\phi-\phi_0=\epsilon P_\nu+\epsilon^2\phi_2+O(\epsilon^3)$ with $L_\nu\phi_2=[\Veff'''(\phi_0)/4\Veff^0]P_\nu^2$ and $\phi_2(1)=0$, and Green's identity as in~\eqref{sm:log-coefficient} gives
\begin{equation}
 \lim_{z\to-1}(1+z)\,\partial_z\phi
 =\epsilon\,\frac{\sin(\pi\nu)}{\pi}+c_2(\nu)\,\epsilon^2+O(\epsilon^3)\,,\qquad
 c_2(\nu)=\frac{\Veff'''(\phi_0)}{8\Veff^0}\int_{-1}^{1}\!\dd z\,P_\nu(-z)\,P_\nu(z)^2\,.
 \label{sm:log-coefficient-asym}
\end{equation}
At an integer, $c_2(n)=-\kappa_n$ for even $n$ and $c_2(n)=0$ for odd $n$.
Setting~\eqref{sm:log-coefficient-asym} to zero near an even integer, where $\sin(\pi\nu)/\pi\simeq\nu-n$, reproduces $\nu-n=\kappa_n\epsilon$, with regular solutions at $\epsilon$ of either sign on the two sides of the integer.

Rescale both couplings by a positive parameter $s$, holding the charges and the coupling ratio fixed.
Then $s\phi_0$ and $\Veff^0$ are constant.
For nonconstant scalar profiles, combining~\eqref{sm:coupling} with~\eqref{sm:averages} gives
\begin{equation}
 \frac{\dd\log\eta}{\dd s}
 =\frac{1}{s\Delta}\int_{r_-}^{r_+}\!\dd r\,
       \frac{(\phi-\phi_0)\partial_\phi\Veff}{b^2}
 =\frac{2}{s\Delta}\int_{r_-}^{r_+}\!\dd r\,|\mathcal R|(\phi')^2>0\,.
 \label{sm:scaling}
\end{equation}
At leading order, $\dd z/\dd\tau=-\Delta(1-z^2)$ and $\phi-\phi_0=\epsilon P_n(z)+O(\epsilon^2)$ give
\begin{equation}
 \int_{r_-}^{r_+}\!\dd r\,|\mathcal R|(\phi')^2
 =\epsilon^2\Delta\int_{-1}^1\!\dd z\,(1-z^2)(P_n')^2+O(\epsilon^3)
 =\frac{2\Delta\epsilon^2n(n+1)}{2n+1}+O(\epsilon^3)\,.
 \label{sm:gradient-leading}
\end{equation}
Since $\nu(\nu+1)\propto s^2$, $\dd s/\dd\nu=s(2\nu+1)/[2\nu(\nu+1)]$; substitution into~\eqref{sm:scaling} gives $\dd\log\eta/\dd\nu=2\epsilon^2+O(\epsilon^3)$.
Integrating with $\nu-n=\kappa_n\epsilon+O(\epsilon^2)$ gives $\eta-1=(2/3)\kappa_n\epsilon^3+O(\epsilon^4)$, yielding the cubic dependence on $\nu-n$ stated in the Letter.

For the excess example $(\alpha_1,\alpha_2)=(\sqrt7,-2\sqrt7)$, $(Q^2,P^2)=(2,1)$, the local coefficient along $(s,-2s)$ is $\kappa_2=6\sqrt6/35$.
The leading excess is therefore $4\sqrt6\,\epsilon^3/35$.
The selected solution has $\phi_+=\phi_-=0.44279366227$ and $\eta=1.02205318802$.
Figure~\ref{sm:fig-tuning} shows it as a path in the $(U,\phi)$ plane.
Autonomous matching and two sided radial shooting with independent horizon series agree within $10^{-12}$ in $\eta$.
A separate collocation solve, seeded by analytic Reissner--Nordstr\"om and $P_2$ profiles, gives $\eta=1.0220531880176$, two scalar nodes, and $\max|b^2H|<2\times10^{-11}$.
Integrating~\eqref{sm:scaling} provides a further consistency check using the same autonomous solutions.

For the asymptotically flat representative with $b_+=5.21819588714$ and seed coordinates $r_+=1$, $r_-=0.35$, the normalized data are
\begin{equation}
 \widehat M\simeq4.167397\,,\qquad
 \phi_\infty\simeq0.939575\,,\qquad
 \widehat\Sigma\simeq-3.134477\,,\qquad
 S_+\simeq85.54421\,,\qquad S_-\simeq1.061268\,.
 \label{sm:excess-data}
\end{equation}
The Kretschmann scalar at the inner horizon is finite, approximately $6.19\times10^4$.
With $k=\Veff(\phi_+)/\Veff^0\simeq2.183278$, the entropy gap bound is $S_+-S_->1.71515\,\Sext$.

\begin{figure}[tbp]
 \centering
 \includegraphics[width=.55\textwidth]{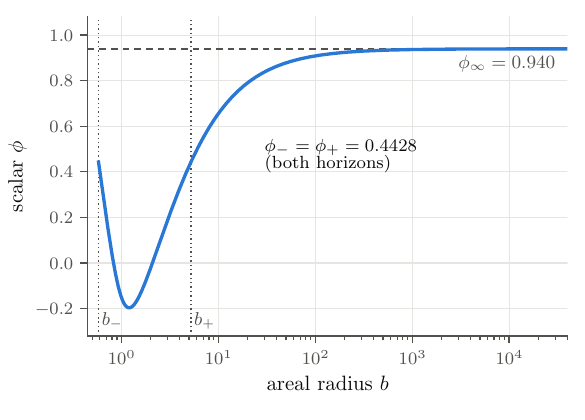}
\caption{\textsf{Scalar profile of the asymmetric excess example in the Letter, plotted against the areal radius.
The two dotted lines mark the inner and outer horizons.
The scalar has two nodes between them and approaches a finite modulus in the asymptotically flat exterior.}}
 \label{sm:fig-excess}
\end{figure}

The asymmetric branch with two nodes reaches an excess of $21.04\%$ at $\epsilon=3$ and a deficit of $7.55\%$ at $\epsilon=-3$.
Separate full interval shooting finds deficits on branches with three nodes at $-\alpha_2/\alpha_1=1.25$, the exact $B_2$ product at $1.5$, and excesses at $2$.
The large $|\epsilon|$ limits and folds are described next.

\section{Branch endpoints and folds}
\label{sm:endpoints}

\begin{figure}[tbp]
 \centering
 \includegraphics[width=.96\textwidth]{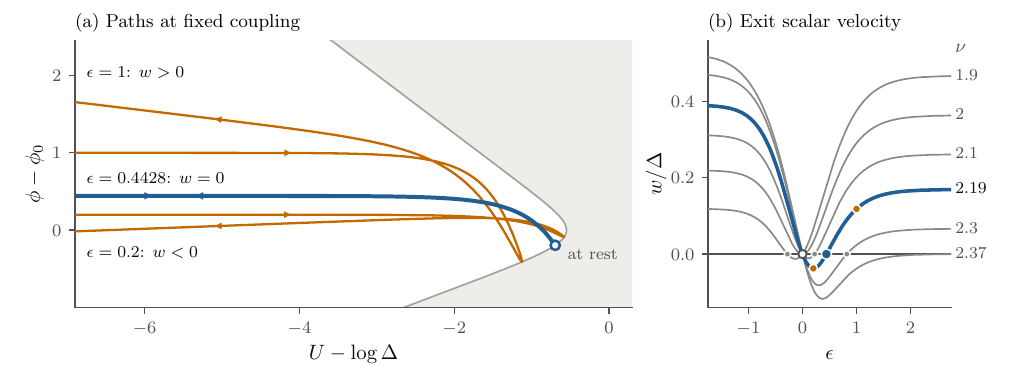}
\caption{\textsf{Tuning the horizon scalar in the excess example of the Letter, $(\alpha_1,\alpha_2)=(\sqrt7,-2\sqrt7)$ with $(Q_1^2,Q_2^2)=(2,1)$.
(a) Interior paths in the $(U,\phi)$ plane; the potential $e^{2U}\Veff$ exceeds $\Delta^2$ in the shaded region.
Each path enters from the outer horizon along $+U$ at $\phi=\phi_0+\epsilon$.
For $\epsilon=0.2$ and $\epsilon=1$ it leaves with scalar velocity $w/\Delta\simeq-0.038$ and $0.118$, and ends at a Kasner singularity.
For $\epsilon\simeq0.4428$ it comes to rest on the boundary of the shaded region, retraces its path, and reaches a regular inner horizon with $\phi_-=\phi_+$.
(b) Exit scalar velocity against $\epsilon$ for couplings $(s,-2s)$ with the same charges, labeled by $\nu$; the thick curve is the theory in (a).
Filled circles on the axis are regular solutions, the open circle is the constant scalar solution, and the two circles off the axis are the singular paths in (a).
At large $|\epsilon|$ each curve approaches its separated reflection value; for $\epsilon\to+\infty$ this value vanishes at the endpoint $\nu_\infty=(\sqrt{33}-1)/2\simeq2.37$.}}
 \label{sm:fig-tuning}
\end{figure}

The autonomous system~\eqref{sm:autonomous} describes motion of $X=(U,\phi)$ in two exponential potentials.
Both horizons lie at $U\to-\infty$, where the exponentials decay and the motion becomes free.
The trajectory enters from the outer horizon with velocity $(\Delta,0)$ at $\phi=\phi_+$, is turned back by the exponentials, and leaves as in~\eqref{sm:kasner-exit} with scalar velocity $w$.
Since $\dd r/\dd\tau=\mathcal R$ gives $r-r_-\propto e^{-2\Delta\tau}$ at the exit, $\phi\simeq-(w/2\Delta)\log(r-r_-)$ there, so $-w/2\Delta$ is the coefficient of the inner horizon logarithm in~\eqref{sm:log-coefficient} and~\eqref{sm:log-coefficient-asym}.
In terms of $U-\log\Delta$, the paths at fixed charges and couplings are labeled by $\epsilon=\phi_+-\phi_0$ alone, and the regular inner horizons are the zeros of $w(\epsilon)$.
Figure~\ref{sm:fig-tuning} shows this tuning for the excess example of the Letter.

\begin{figure}[tbp]
 \centering
 \includegraphics[width=.96\textwidth]{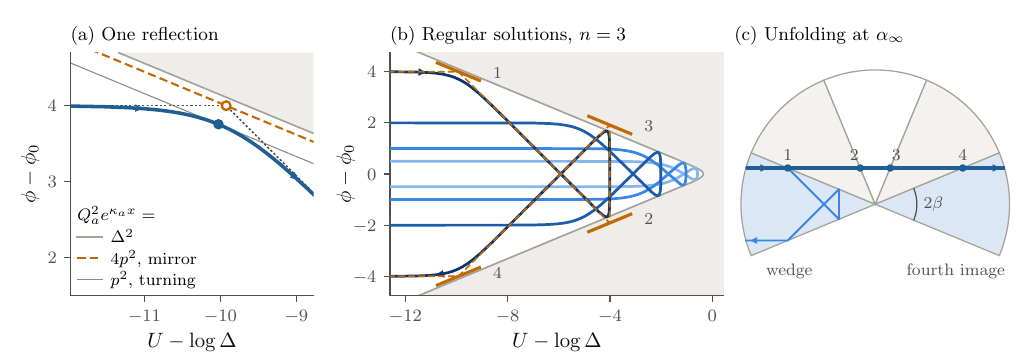}
\caption{\textsf{Separated reflections in the symmetric model with unit charges.
(a) Exact path~\eqref{sm:liouville} in a single exponential, entering with velocity $(\Delta,0)$ and normal speed $p$.
It turns where the exponential equals $p^2$ (filled circle); its asymptotes (dotted) meet on the mirror line, where the exponential equals $4p^2$ (open circle).
(b) Regular solutions with three nodes at $\epsilon=0.5$, $1$, $2$, and $4$ (light to dark), each at its own tuned coupling; the shaded region is drawn for the endpoint coupling.
The dashed path consists of straight segments joined at the displaced mirrors (bars) of the four numbered reflections, at the endpoint coupling $\alpha_\infty=2\cot(\pi/8)$; panel (a) is its first reflection.
At $\epsilon=4$ the tuned coupling has $\nu-\nu_\infty\simeq3.3\times10^{-7}$.
(c) The same four reflections unfolded, with the walls drawn as lines through one apex.
Reflecting the wedge, of opening angle $2\beta=2\arctan(2/\alpha)$, instead of the path gives a straight line through four images of the wedge; the folded path is shown in the wedge itself.
The exit is antiparallel to the entry because the fourth image is the wedge turned by $\pi$, that is, $4\times2\beta=\pi$.}}
 \label{sm:fig-reflections}
\end{figure}

The separated exponential limit at large $|\epsilon|$ consists of motion with speed $\Delta$ between widely separated reflections, each dominated by one exponential.
Figure~\ref{sm:fig-reflections}(b) shows regular solutions approaching this limit as $\epsilon$ increases.
We first calculate the coupling and entropy product for a fixed reflection sequence, then compare these limiting predictions with smooth numerical solutions.
Set $\kappa_a=\sqrt{4+\alpha_a^2}$, $\boldsymbol n_a=(2,\alpha_a)/\kappa_a$, and $x=\boldsymbol n_a\cdot X$.
For incoming normal speed $p>0$, the single exponential problem has the exact solution
\begin{equation}
 \dot x^2+Q_a^2e^{\kappa_ax}=p^2\,,\qquad
 Q_a^2e^{\kappa_ax}=p^2\operatorname{sech}^2\!\left[\frac{\kappa_ap}{2}(\tau-\tau_*)\right]\,,\qquad
 x\simeq\frac1{\kappa_a}\log\frac{4p^2}{Q_a^2}-p|\tau-\tau_*|\,.
 \label{sm:liouville}
\end{equation}
Here, $\tau_*$ is the reflection time, when the normal velocity vanishes.
The incoming and outgoing asymptotes meet at the displaced mirror $x_\text{mirror}=\kappa_a^{-1}\log(4p^2/Q_a^2)$.
Thus each exponential acts on asymptotic trajectories as a mirror with a speed dependent displacement, as shown in Figure~\ref{sm:fig-reflections}(a).

The trajectory enters along $+U$, with $U=\tau\Delta+u_0$ and $\phi=\phi_+$.
A regular inner horizon requires exit along $-U$, with $U=-\tau\Delta+u_1$ and $\dot\phi=0$.
The reflections alternate between the two walls; each intervening flight crosses $\phi_0$, so $n$ scalar nodes correspond to $n+1$ reflections.
Writing $\beta_a=\arctan(2/|\alpha_a|)$, composition of these reflections gives
\begin{equation}
 (n+1+k)\beta_1+(n+1-k)\beta_2=\pi\,,\qquad
 k=\begin{cases}0,&n\text{ odd},\\ \pm1,&n\text{ even},\ \epsilon\to\pm\infty\end{cases}\,.
 \label{sm:endpoint-angle}
\end{equation}
For even $n$, $k=+1$ means that the first reflection is from the $Q_1^2e^{2U+\alpha_1\phi}$ wall, as $\epsilon\to+\infty$; $k=-1$ exchanges the walls.
This fixes the limiting coupling for the specified reflection sequence.
Figure~\ref{sm:fig-reflections}(c) shows the condition for symmetric couplings after unfolding.
Reflecting the wedge between the walls, instead of the trajectory, turns the path into a straight line through successive images of the wedge, each subtending $2\beta$ at the apex.
The exit is antiparallel to the entry when the image reached after $n+1$ reflections is the wedge turned by $\pi$, that is, when $n+1$ images fill a half plane.
At a nearby coupling with the same reflection sequence, the velocity instead turns through $\Theta=(n+1+k)\beta_1+(n+1-k)\beta_2$, so that $w/\Delta\to(-1)^n\operatorname{sgn}(\epsilon)\sin(\Theta-\pi)$ as $|\epsilon|\to\infty$.
These are the limiting values in Figure~\ref{sm:fig-tuning}(b); the endpoint coupling is the one at which they vanish.

Let $\hat p_j$ be the normal component of the unit incoming velocity at reflection $j$, with $\kappa_j$, $Q_j$, and $\boldsymbol n_j$ belonging to the wall encountered.
The mirror displacement at that encounter is $m_j=\kappa_j^{-1}\log(4\hat p_j^2\Delta^2/Q_j^2)$, since the actual incoming normal speed is $\hat p_j\Delta$.
Write each free asymptote as $X=\widehat{\boldsymbol v}\tau\Delta+\boldsymbol a$.
Reflection changes its longitudinal intercept by $\widehat{\boldsymbol v}_\text{out}\cdot\boldsymbol a_\text{out}-\widehat{\boldsymbol v}_\text{in}\cdot\boldsymbol a_\text{in}=-2\hat p_jm_j$.
Summing over the reflections, whose initial and final velocities are $(1,0)$ and $(-1,0)$, gives
\begin{equation}
 u_0+u_1=2\sum_{j=1}^{n+1}\hat p_jm_j\,,\qquad
 b_+b_-=4\Delta^2e^{-(u_0+u_1)}\,.
 \label{sm:mirror-intercepts}
\end{equation}
The second identity follows by integrating $\dd r/\dd\tau=\mathcal R$ and reconstructing $b^2=-\mathcal R e^{-2U}$ at the horizons.
Thus the affine displacements determine the entropy product:
\begin{equation}
 \log\eta_\infty=-4\sum_{j=1}^{n+1}\frac{\hat p_j}{\kappa_j}\log\frac{\hat p_j^2\Veff^0}{Q_j^2}\,,\qquad
 \sum_{j=1}^{n+1}\hat p_j\boldsymbol n_j=(1,0)\,.
 \label{sm:endpoint-product}
\end{equation}
The $U$ component of the impulse identity gives $\sum_j\hat p_j/\kappa_j=1/2$.
Substituting $m_j$ into~\eqref{sm:mirror-intercepts} therefore cancels $4\log 2\Delta$ from $\log[(b_+b_-)^2]$.
The scalar component fixes the relative weights of the two walls and cancels the charge ratio after normalization by $(\Veff^0)^2$.
For symmetric couplings, $\beta=\pi/(2n+2)$, $\kappa=2/\sin\beta$, and $\hat p_j=\sin[(2j-1)\beta]$, yielding
\begin{equation}
 \nu_\infty(\nu_\infty+1)=2\cot^2\beta\,,\qquad
 \eta_\infty(n)=\frac14\exp\!\left[-4\sin\beta\sum_{j=1}^{n+1}\hat p_j\log\hat p_j\right]\,.
 \label{sm:symmetric-endpoint}
\end{equation}
The dashed path in Figure~\ref{sm:fig-reflections}(b) is built from these mirrors for $n=3$.
As $n\to\infty$, the sum becomes a midpoint integral, with $\int_0^\pi\!\dd x\,\sin x\log\sin x=2\log2-2$, giving $\eta_\infty\to e^4/64$.
The limiting couplings $\nu_\infty(13)=12.06144755$ and $\nu_\infty(14)=12.96463003$ reproduce Davydov's numerical thresholds for closing gaps and the onset of overlapping symmetric windows~\cite{Davydov:2017zbs}.

\noindent\textit{Asymmetric couplings.}\quad
Set $(\alpha_1,\alpha_2)=(s,-\zeta s)$ with $\zeta>1$.
For two nodes, the $\epsilon>0$ itinerary is $1,2,1$, with normal speeds $(\sin\beta_1,1,\sin\beta_1)$; exchanging the walls gives the $\epsilon<0$ end.
Equations~\eqref{sm:endpoint-angle} and~\eqref{sm:endpoint-product} reduce to
\begin{equation}
 \begin{gathered}
 s_{\infty,+}^2=\frac{4(\zeta+2)}{\zeta}\,,\qquad
 \eta_{\infty,+}=\left(\frac{2^\zeta}{\zeta+1}\right)^{2/(\zeta+1)}\,,\\
 s_{\infty,-}^2=\frac{4(2\zeta+1)}{\zeta^2}\,,\qquad
 \eta_{\infty,-}=\left(\frac{2\zeta^\zeta}{(\zeta+1)^\zeta}\right)^{2/(\zeta+1)}\,.
 \end{gathered}
 \label{sm:two-node-endpoints}
\end{equation}
The signs here label $\epsilon\to\pm\infty$.
At $\zeta=2$, the endpoints are $(\nu,\eta)=((\sqrt{21}-1)/2,(8/9)^{2/3})$ and $((\sqrt{33}-1)/2,(4/3)^{2/3})$.
Moreover, $\partial_\zeta\log\eta_{\infty,+}=2[\log(2\zeta+2)-1]/(\zeta+1)^2>0$, so this endpoint product increases from $1$ toward $4$; at $\zeta=3$ it equals $\sqrt2$.
For three nodes, the angular condition reduces to $\zeta s^2-2(\zeta+1)s-4=0$.
At $\zeta=2$ it gives $s=(3+\sqrt{17})/2$ and $\nu_\infty\simeq3.09647862$, accounting for the three node endpoint in Figure~1(b) of the Letter.
It gives $s=4$, $\nu=3$, and $\eta_\infty=1$ at the $B_2$ ratio $\zeta=3/2$.
The $A_1\oplus A_1$ locus $\alpha_1|\alpha_2|=4$ and the $A_2$, $B_2$, and $G_2$ Toda points also satisfy the endpoint condition at their integer indices~\cite{Davydov:2017zbs,Zadora:2017qeo}.
Their families of varying $\epsilon$ have constant $(\nu,\eta)=(n,1)$ and therefore collapse to points in the Letter's plot.

\noindent\textit{Folds and checks.}\quad
Away from the constant scalar solution,~\eqref{sm:scaling} gives $\dd\log\eta/\dd\epsilon=K(\epsilon)\,\dd\nu/\dd\epsilon$ with $K>0$ along a branch parametrized smoothly by $\epsilon$.
Thus a turning point of $\nu$ is also a turning point of $\eta$.
For $\zeta=3$ and four nodes, the $\epsilon<0$ side turns at $(\epsilon,\nu,\eta)\simeq(-0.37943621,3.77351998,0.99108426)$ before approaching $(\nu_\infty,\eta_\infty)\simeq(3.91447667,1.12522751)$.
A turn is required on any smooth continuation from $(\nu,\eta)=(4,1)$ to the stated endpoint.
With $A=-\epsilon>0$, the local crossing coefficient is positive, so both $\nu$ and $\eta$ initially decrease from $(4,1)$.
Since $\eta_\infty>1$, the product must subsequently increase, and the positive sign identity forces the coupling to increase with it.
This argument establishes the need for at least one turn conditional on that connection; continuation determines its location and the observed single fold.
At fixed $\nu=3.85$ and $(Q_1^2,Q_2^2)=(3,1)$, independent shooting gives two four node solutions,
\begin{equation}
 (\epsilon,\eta)\simeq(-0.14469704,0.99820684)\,,\qquad
 (\epsilon,\eta)\simeq(-0.85970298,1.03417346)\,.
 \label{sm:fold-pair}
\end{equation}
Full interval integration gives inner scalar velocity residuals below $2\times10^{-15}$ and energy residuals below $6\times10^{-14}$.
The interior comparison fixes charges and couplings.
The exterior translation orbit also permits the two solutions to be compared at the common modulus $\phi_\infty=-1$, as shown in Table~\ref{sm:table-fold-modulus}.
For this check, set $U=\tfrac12\log a^2$ outside the outer horizon and integrate $\ddot U=e^{2U}\Veff$ and $\ddot\phi=e^{2U}\partial_\phi\Veff/2$ with $\Delta=1$ and $U\sim\tau$ at that horizon.
If $\tau=s_*$ is the point where $\phi=-1$, the normalized data follow from $b_+=2e^{-s_*}$, $c=e^{-U(s_*)}$, $\widehat M=c\dot U(s_*)$, and $\widehat\Sigma=-c\dot\phi(s_*)$; these are the orbit observables~\eqref{sm:orbit-observables}.
Starting at $\tau=-30$ with relative tolerance $10^{-13}$, and refining to $\tau=-32$ with tolerance $3\times10^{-14}$, leaves the table unchanged at the displayed precision.
Independent electric potential integrals give a Smarr residual below $2\times10^{-13}$; the refined asymptotic constraint $\widehat M^2+\widehat\Sigma^2-\Veff(\phi_\infty)-c^2$ is below $10^{-9}$ in absolute value.
The comparison fixes the asymptotic vacuum as well as the charges and theory.
The masses and temperatures differ at this modulus, so thermodynamic coexistence in a specified ensemble requires a separate analysis.

\begin{table}[tbp]
 \centering
 \begin{tabular}{lrrr}
 \toprule
 Branch & $\phi_+$ & $\eta$ & $\widehat M$\\
 \midrule
 Deficit & $-0.14469704$ & $0.99820684$ & $41.58209437$\\
 Excess & $-0.85970298$ & $1.03417346$ & $365.79681168$\\
 \bottomrule
 \end{tabular}
\caption{\textsf{Two regular four node black holes at the same asymptotic modulus $\phi_\infty=-1$, charges $(Q_1^2,Q_2^2)=(3,1)$, and coupling $\nu=3.85$ along $(\alpha_1,\alpha_2)=(s,-3s)$.
The mass uses unit lapse at infinity.}}
 \label{sm:table-fold-modulus}
\end{table}
Direct integration at $|\epsilon|=8$ reproduces the two and three node endpoints at $\zeta=2$ and the four node endpoints at $\zeta=3$ within $6\times10^{-9}$ in $\eta$.
The independent calculation starts at normalized autonomous time $u=-25$, uses DOP853 with relative tolerance $10^{-13}$ and absolute tolerance $10^{-15}$, and checks scalar node counts and the energy constraint.

The reflection formulas determine the separated exponential limit exactly.
Their transfer to a smooth family requires convergence of the affine intercepts in~\eqref{sm:mirror-intercepts}, as well as of the velocities.
For a fixed finite sequence of uniformly nongrazing reflections, a sufficient separation estimate is exponential suppression of the inactive wall, flight times growing at most linearly with $A=|\epsilon|$, and $A|s(A)-s_\infty|\to0$.
Under these assumptions the smooth single reflection maps accumulate only a polynomial factor multiplying the exponentially small overlap error, while the coupling displacement gives a vanishing intercept correction.
We use the resulting reflection values as endpoint predictions in the separated reflection approximation; the numerical convergence checks above test those predictions for the continued branches.
A global continuation and separation estimate for every branch has not been established.
For a monotone continuation the two endpoints bound the coupling window; when folds occur, its boundaries are determined by the global coupling extrema.
In particular, the folded branch reaches its minimum coupling at finite $\epsilon$.
For an asymptotically flat sequence approaching an endpoint at $|\epsilon|\to\infty$, the outer constraint gives $S_+>\pi\Veff(\phi_+)\to\infty$, while finite $\eta_\infty\Sext^2$ requires $S_-\to0$.
The finite point in the $(\nu,\eta)$ plane therefore represents a singular limit of regular solutions at finite $\epsilon$.
The symmetric limiting deficit and the two node limiting excess above characterize their respective endpoint sequences.

\noindent\textit{Returns through an integer.}\quad
For $(\alpha_1,\alpha_2)=(s,-\zeta s)$ with $\zeta>1$ and $(Q_1^2,Q_2^2)=(\zeta,1)$, one has $\phi_0=0$ and $\Veff'''(0)=\zeta s^3(1-\zeta^2)<0$, so $\kappa_4>0$.
The $\epsilon>0$ and $\epsilon<0$ four node branches therefore leave $\nu=4$ upward and downward, respectively.
At $\nu=4$, $s^2=40/\zeta$ and the wall angles satisfy $\tan\beta_1=\sqrt{\zeta/10}$ and $\tan\beta_2=1/\sqrt{10\zeta}$.
The endpoint conditions $3\beta_1+2\beta_2=\pi/2$ and $2\beta_1+3\beta_2=\pi/2$ reduce to
\begin{equation}
 28\zeta^2-43\zeta+10=0\,,\qquad
 10\zeta^2-43\zeta+28=0\,,
 \label{sm:integer-thresholds}
\end{equation}
whose roots with $\zeta>1$ are $5/4$ and $7/2$, respectively.
For $1<\zeta<5/4$ the $\epsilon>0$ endpoint lies below $\nu=4$, while for $\zeta>7/2$ the $\epsilon<0$ endpoint lies above it.
Any continuous branch connecting the corresponding Reissner--Nordstr\"om point and branch endpoint must therefore fold and return through $\nu=4$ at finite $\epsilon$.

\begin{figure}[tbp]
 \centering
 \includegraphics[width=.62\textwidth]{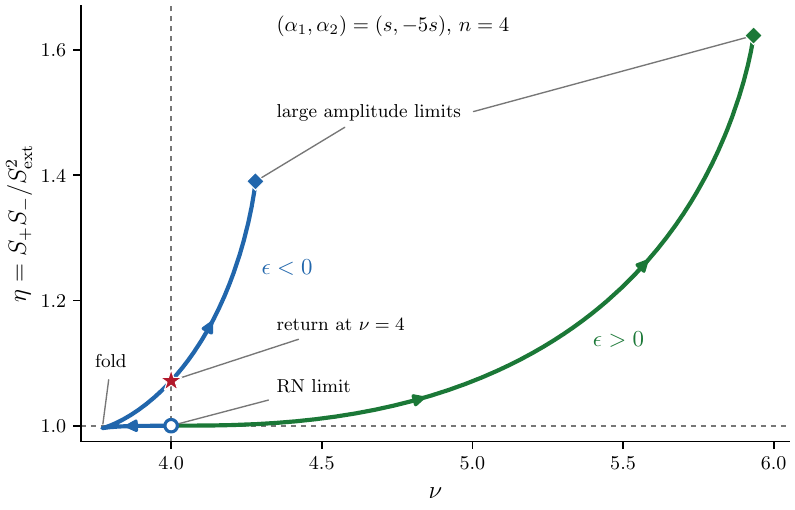}
\caption{\textsf{Both four node legs for couplings $(\alpha_1,\alpha_2)=(s,-5s)$ and $(Q_1^2,Q_2^2)=(5,1)$.
Starting at the open Reissner--Nordstr\"om point, increasing $|\epsilon|$ follows the arrows.
The $\epsilon<0$ leg (blue) passes through a fold and returns to $\nu=4$ with $\eta=1.071532$ (star); the $\epsilon>0$ leg (green) rises monotonically.
Diamonds are the separated reflection predictions for the branch endpoints, $(\nu_\infty,\eta_\infty)\simeq(4.279977,1.390321)$ and $(5.933187,1.623180)$.}}
 \label{sm:fig-integer-return}
\end{figure}

Smooth equation continuation realizes both possibilities.
At $\zeta=1.1$ the return has $(\epsilon,\eta)\simeq(0.33961006,0.99886327)$; at $\zeta=5$ it has $(-0.62897562,1.07153225)$.
Both have four scalar nodes, with full interval energy residuals below $3\times10^{-14}$ and inner scalar velocity below $5\times10^{-15}$.
An independently implemented integration in unscaled $U$ reproduces the products within $2\times10^{-11}$.
The regular horizon construction and the orbit map apply to these solutions at finite $\epsilon$ and supply asymptotically flat families.
Figure~\ref{sm:fig-integer-return} displays both four node legs at $\zeta=5$.
The $\epsilon>0$ leg rises monotonically from $\nu=4$, without a fold, to its endpoint $(\nu_\infty,\eta_\infty)\simeq(5.933187,1.623180)$.
On $4<\nu<4.28$ the two legs give two excess solutions with different products; over the sampled range $|\epsilon|\le10$ the curves do not cross.

The return product measures the accumulated scalar gradient response along the excursion.
For the $\epsilon<0$ leg, let $s_4=\sqrt8$ and let $s_\text{min}$ denote its minimum coupling.
Writing~\eqref{sm:scaling} as $\dd\log\eta=G\,\dd\log s$ gives
\begin{equation}
 \log\eta_\text{return}
 =\int_{s_\text{min}}^{s_4}\!\frac{\dd s}{s}\,
 (G_\text{back}-G_\text{out})\,,\qquad
 G=\frac2\Delta\int_{r_-}^{r_+}\!\dd r\,|\mathcal R|(\phi')^2\,.
 \label{sm:integer-return-action}
\end{equation}
The sampled gradient integral is larger on the returning leg, giving the excess.
An excursion above the integer has the opposite orientation.
This explains the numerical signs; positivity of $G$ alone does not fix a return product, and the local sign rule near $\nu=n$ does not extend across a fold.

\noindent\textit{Toda couplings and supergravity.}\quad
At a rank two Toda coupling the exact solutions cover all finite regular outer horizon scalar data and connect them continuously to Reissner--Nordstr\"om through regular interiors~\cite{Davydov:2017zbs,Zadora:2017qeo,Lu:2026lkv}.
Constancy of the product on each connected regular family consequently fixes $\eta=1$ throughout these families.
Whether an open interval of $\epsilon$ with regular horizons at fixed coupling requires a Toda theory remains conjectural.
The relevant condition is the full coupling pair: a Toda coupling ratio at another value of $\nu$ need not define a Toda theory.
The standard STU sectors have couplings $(\alpha_1,\alpha_2)=(2,-2)$, $(2\sqrt3,-2/\sqrt3)$, and $(2\sqrt3,-2\sqrt3)$~\cite{Lu:2026lkv}; the first two lie on the $A_1\oplus A_1$ locus $\alpha_1\alpha_2=-4$, and the third is the $A_2$ point.
The ungauged $\mathcal N=2$ models with monomial prepotential $\mathcal F=-i(X^0)^p(X^1)^{2-p}/4$, $0<p<2$, provide a continuous class: their purely electric sector with vanishing axion has $\alpha_1\alpha_2=-4$ and hence $\eta=1$~\cite{Anabalon:2020pez}.
This includes non-BPS black holes in those supersymmetric parent theories.

Supersymmetry of a parent theory by itself permits more general couplings.
For example, a classical $\mathcal N=1$ completion uses one neutral chiral field $Z=\phi+i\chi$, two Abelian vectors and
\begin{equation}
 \mathcal K=-\tfrac12(Z-\bar Z)^2\,,\qquad
 W=0\,,\qquad f_{ab}=2\delta_{ab}e^{-\alpha_a Z}\,.
 \label{sm:n1-parent}
\end{equation}
In the conventional normalization with Einstein term $R/2$, $\mathcal K_{Z\bar Z}=1$, and no Fayet--Iliopoulos terms, the scalar potential vanishes~\cite{Jockers:2004yj}.
Multiplying that Lagrangian by two and setting $\chi=0$ gives the Letter's action for arbitrary couplings.
The discarded scalar equation holds in this purely electric sector: $\partial_\chi\operatorname{Re}f_{ab}|_{\chi=0}=0$ and every $F^a\wedge F^a$ vanishes.
The gauge kinetic matrix is positive in an open strip about $\chi=0$.
Thus the returning non-BPS black holes also solve this $\mathcal N=1$ theory, whose axion partner is present but has zero profile.
This construction establishes a classical supersymmetric parent; it supplies neither an $\mathcal N=2$ embedding nor a string compactification.

\section{Numerical methods, checks, and coupling windows}
\label{sm:numerics}

The radial constructions use $r_+=1$ and $r_-=0.35$.
The symmetric example starts from the Reissner--Nordstr\"om profile $b_0$ in the End Matter and scalar seed $0.2P_3(z)$.
These profiles initialize the nonlinear solve; they are not solutions at nonzero $\epsilon$.
At either horizon write $s=r-r_h$, $\mathcal R=d_hs+s^2$, $b=\sum b_ns^n$, and $\phi=\sum q_ns^n$, with $d_h=\mathcal R'(r_h)$.
The general recursion is
\begin{equation}
 d_h(n+1)^2q_{n+1}+n(n+1)q_n
 =\left[\frac{\partial_\phi\Veff}{2b^2}\right]_n\,,\qquad
 (n+2)(n+1)b_{n+2}=-[b(\phi')^2]_n\,,\qquad
 b_1=\frac{b_0-\Veff(q_0)/b_0}{d_h}\,.
 \label{sm:series}
\end{equation}
Brackets denote the coefficient of the indicated power of $s$.
Series truncated at different orders supply data a small distance from each horizon into the integration interval.
The omitted endpoint intervals are covered by these series; dense numerical output is not extrapolated to the horizons.

Collocation evolves $(b,b',\phi,j)$ with $j=\mathcal R\phi'$ and matches all four fields to both endpoint series.
At fixed $\epsilon$ in the symmetric model, the unknown parameters are $(\alpha,b_-,b'_-,\phi_-)$; at fixed coupling, $\phi_+$ replaces $\alpha$.
An independent shooting construction imposes both horizon constraints, matches $b,\phi,\phi'$ at an interior point, and checks the mismatch in $b'$, which is not imposed in the matching.
The asymmetric implementation substitutes its potential into the recurrence and tunes the coupling parameter $s$ or the horizon scalar $\phi_+$.

For $\epsilon=0.2$ and $b_+=2.5b_+^{(0)}$, where the Reissner--Nordstr\"om reference radius satisfies $b_+^{(0)}(b_+^{(0)}-0.65)=2$, the independent methods give
\begin{equation}
 \alpha=4.89376036208404\,,\qquad
 \nu=2.9963474141956\,,\qquad
 \eta=0.999859966473997\,.
 \label{sm:deficit-point}
\end{equation}
The superscript labels the constant scalar reference; its horizons are distinct.
The scaled differential equation residual sampled away from the solver mesh is below $8.1\times10^{-10}$, and the reference solution has $\max|b^2H|<6.7\times10^{-13}$.
The Kretschmann scalar at the inner horizon is approximately $1.00482\times10^5$, large but finite.
Repeating the solve at fixed $\alpha$ for three outer radii changes $\phi_+$ by less than $2.4\times10^{-12}$ and $\eta$ by less than $3.0\times10^{-13}$.

\begin{table}[tbp]
 \centering
 \begin{tabular}{lcccc}
 \toprule
 Method & Series order & Cutoff & $\eta$ & $\max|b^2H|$\\
 \midrule
 Collocation & 4 & $3\times10^{-5}$ & 0.999859966390 & $3.5\times10^{-10}$\\
 Collocation & 8 & $10^{-4}$ & 0.999859966474 & $6.6\times10^{-13}$\\
 Two sided shooting & 3 & $10^{-5}$ & 0.999859971346 & $4.4\times10^{-9}$\\
 Two sided shooting & 3 & $10^{-7}$ & 0.999859966474 & $1.8\times10^{-12}$\\
 \bottomrule
 \end{tabular}
\caption{\textsf{Independent endpoint refinements at $\epsilon=0.2$.
The collocation rows with series orders four and eight use normalized residual tolerances $10^{-8}$ and $10^{-10}$, respectively.
Shooting uses relative integration tolerance $2\times10^{-13}$ and absolute tolerance one hundred times smaller.
The spread between results measures empirical convergence.}}
 \label{sm:table-convergence}
\end{table}

Between the horizons, define $U=\tfrac12\log(-a^2)$.
Continuation to finite $\epsilon$ uses the autonomous system
\begin{equation}
 \ddot U=-e^{2U}\Veff\,,\qquad
 \ddot\phi=-\tfrac12e^{2U}\partial_\phi\Veff\,,\qquad
 \dot U^2+\dot\phi^2+e^{2U}\Veff=\Delta^2\,,
 \qquad \frac{\dd r}{\dd\tau}=\mathcal R\,.
 \label{sm:autonomous}
\end{equation}
For the symmetric potential, use $u=\tau\Delta$ up to a translation and $w=U-\log\Delta+\frac12\log\Veff(\epsilon)$, fixing the translation by $w-u\to0$ as $u\to-\infty$ at the outer horizon.
The constraint becomes
\begin{equation}
 w_{,u}^2+\phi_{,u}^2+
 e^{2w}\frac{\cosh(\alpha\phi)}{\cosh(\alpha\epsilon)}=1\,.
 \label{sm:scaled}
\end{equation}
The potential ratio is evaluated logarithmically.
DOP853 integration starts at $u=-18$ in this outer asymptotic normalization, with regular series data, and stops at the unique maximum of $w$.
The location of that maximum is determined by the solution and is not fixed to $u=0$.
Matching imposes $\phi=0$ for odd modes and $\phi_{,u}=0$ for even modes.
Counting zeros on the reflected interval verifies the scalar node number.
Davydov's convention is $\psi=2(\phi-\phi_0)$ and $a=\alpha/2$, which leaves $\nu$ and these node counts unchanged~\cite{Davydov:2017zbs}.

\begin{figure}[tbp]
 \centering
 \includegraphics[width=.9\textwidth]{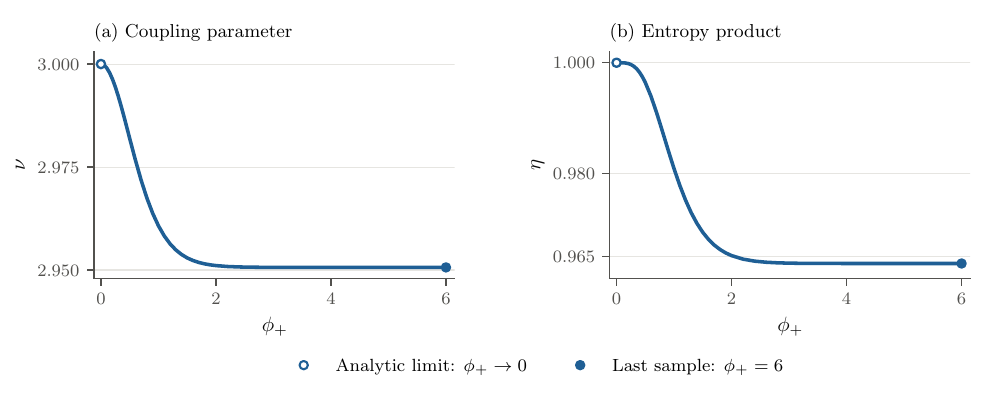}
\caption{\textsf{Dependence of the symmetric branch with three scalar nodes on the horizon scalar.
The coupling index $\nu$ and product ratio $\eta$ are shown against the selected horizon scalar $\phi_+$.
Open markers are the analytic $\phi_+\to0$ limits and filled markers are the final samples at $\phi_+=6$.
Each point represents an orbit at fixed coupling; moving along the curve changes the theory.
The combined $\eta(\nu)$ curves appear in the Letter.}}
 \label{sm:fig-branches}
\end{figure}

Each branch is sampled at 45 nonzero values from $\epsilon=0.05$ to $6$.
Repeating the final sample with starting points $u=-16$ and $-20$ and tighter tolerances gives a maximum spread in $\eta$ of $1.9\times10^{-9}$.
For the branch with three scalar nodes, independent radial shooting agrees over the overlapping range $0.05\le\epsilon\le0.35$ within $1.6\times10^{-12}$ in $\alpha$ and $4.6\times10^{-13}$ in $\eta$.
The full interval comparison at $\epsilon=0.2$ is shown in Table~\ref{sm:table-convergence}.
The radial methods have not independently verified the samples at large $\epsilon$.
The separated reflection endpoint predictions provide an independent check on the asymptotic end of the autonomous continuation.
Runs with failed integration, missing matching events, incomplete intervals, or nonfinite profiles are rejected; constraint residuals, matching residuals, and node counts are recorded.

\begin{table}[tbp]
 \centering
 \begin{tabular}{crrrrrr}
 \toprule
 $n$ & $\nu$ at $\epsilon=6$ & $n-\nu$ &
 \multicolumn{3}{c}{Deficit $100(1-\eta)$} & $\eta_\infty$\\
 &&& $\epsilon=1$ & $\epsilon=2$ & $\epsilon=6$ & reflection\\
 \midrule
 3 & 2.950631 & 0.049369 & 1.9178 & 3.4901 & 3.6330 & 0.96366977\\
 4 & 3.881127 & 0.118873 & 3.9815 & 6.1719 & 6.3207 & 0.93679276\\
 5 & 4.801548 & 0.198452 & 5.6737 & 8.0295 & 8.1676 & 0.91832430\\
 6 & 5.716216 & 0.283784 & 6.9718 & 9.3353 & 9.4633 & 0.90536670\\
 7 & 6.627292 & 0.372708 & 7.9587 & 10.2819 & 10.4021 & 0.89597890\\
 \bottomrule
 \end{tabular}
\caption{\textsf{Sampled coupling ranges and deficits.
The final column gives the separated reflection prediction~\eqref{sm:symmetric-endpoint}.
The $\epsilon=6$ deficits agree with its corresponding percentage deficits at the displayed precision; independent integration gives differences below $1.7\times10^{-9}$ in $\eta$.}}
 \label{sm:table-windows}
\end{table}

Solving for the horizon scalar at fixed $\nu=2.99$ gives $\phi_+=0.34680368$ and $\eta=0.99893468$.
The fixed coupling scans distinguish the selected regular roots from the constant scalar solution, but finite sampling cannot exclude tangent roots, closely spaced pairs, solutions at larger $\epsilon$, or solutions without reflection symmetry.
The widths in Table~\ref{sm:table-windows} grow with node number; we have not scanned the entropy products in the overlapping windows at larger $n$ reported by Davydov.

At large $\epsilon$, the horizon areas differ by many orders of magnitude.
For $n=3$, $k=\cosh(\alpha\epsilon)$ grows from $5.76$ at $\epsilon=0.5$ to $63.42$ at $1$, $7.82\times10^3$ at $2$, and $1.91\times10^{12}$ at $6$.
Every asymptotically flat member obeys $S_+/\Sext>k$ and $S_-/\Sext<\eta/k$.
The deficit is already $1.9178\%$ at $\epsilon=1$; it is not confined to the largest hierarchy at the end of the scan.
The autonomous formulation resolves these narrow radial layers.
As exact controls, the symmetric $A_2$ and asymmetric $B_2$ Toda solutions have $\eta=1$~\cite{Zadora:2017qeo}.
The $A_2$ control returns $|\eta-1|<1.6\times10^{-13}$ at $\epsilon=0.2,1,3,6$.

\section{Exterior families and extremal comparison}
\label{sm:exteriors}

The exact orbit map and asymptotic normalization are given in the End Matter.
In those variables the exterior metric is
\begin{equation}
 \dd s_\lambda^2=-c_\lambda^2\frac{\mathcal R(r)}{b_\lambda(r)^2}
                  \dd\widehat t^2
              +\frac{b_\lambda(r)^2}{\mathcal R(r)}\dd r^2
              +b_\lambda(r)^2\dd\Omega_2^2\,.
 \label{sm:orbit-metric}
\end{equation}
The root separation in asymptotically normalized radial coordinates is $2c_\lambda \Delta$, so fixing $\Delta$ in the seed coordinates does not fix the physical nonextremality along the orbit.
The transformation is the radial translation symmetry of the autonomous dynamics~\cite{Ferrara:1997tw,Galli:2011fq}.
On a nonconstant exterior, convexity of $\Veff$ and the scalar equation imply $\operatorname{sgn}\phi'=\operatorname{sgn}(\phi_+-\phi_0)$: the flux $\mathcal R\phi'$ vanishes at the regular outer horizon and then grows with that sign.
The orbit map samples this strictly monotone exterior scalar profile at infinity, so $\phi_\infty$ varies strictly along each such orbit.
Fixing the charges, couplings, and $\phi_\infty$ therefore selects at most one member of that orbit, and hence one mass; it does not classify other regular branches at the same boundary data.

\noindent\textit{A common vacuum.}\quad
The classical field equations admit the exact transformation
\begin{equation}
 \widetilde\phi=\phi-\vartheta\,,\qquad
 \widetilde F^a=e^{-\alpha_a\vartheta/2}F^a\,,\qquad
 \widetilde Q_a=e^{\alpha_a\vartheta/2}Q_a\,,\qquad
 \widetilde g_{\mu\nu}=g_{\mu\nu}\,.
 \label{sm:fixed-vacuum-shift}
\end{equation}
Indeed, $e^{-\alpha_a\widetilde\phi}(\widetilde F^a)^2=e^{-\alpha_a\phi}(F^a)^2$, while the conserved electric flux is weighted by $e^{-\alpha_a\widetilde\phi}$.
Choosing $\vartheta=\phi_\infty-\bar\phi$ puts every orbit member in the same asymptotic vacuum $\widetilde\phi_\infty=\bar\phi$, at unchanged couplings.
This is a classical field symmetry, not a claim that arbitrary continuous rescalings preserve a quantum charge lattice.
The metric, $M$, $\Sigma$, $T_h$, and $S_h$ are unchanged.
Both charge magnitudes, and hence their ratio, generally vary; this is not a family at fixed charges.
Since $\widetilde\Veff(\widetilde\phi)=\Veff(\widetilde\phi+\vartheta)$, its minimum, $\Sext$, and $\eta$ are unchanged.
The dressed charges
\begin{equation}
 \mathcal Q_a=|Q_a|e^{\alpha_a\phi_\infty/2}
     =|\widetilde Q_a|e^{\alpha_a\bar\phi/2}
 \label{sm:dressed-charges}
\end{equation}
are convenient invariant measures of the electromagnetic scales.

At fixed $\bar\phi$, the constant scalar Reissner--Nordstr\"om sector satisfies the charge constraint $\partial_\phi\Veff(\bar\phi)=0$.
A selected nonconstant orbit instead supplies at most one mass at given charges and $\bar\phi$, when that modulus lies in its exterior range.
Where the regularity condition is transverse, these are locally codimension one sets in the space of solutions with a regular outer horizon parametrized by $(M,Q_1,Q_2)$.
The analytic branches near integer $\nu$ and the numerical branches studied here establish such tuning locally; they do not classify every regular branch at arbitrary couplings.

\noindent\textit{Missing the regular inner horizon.}\quad
Davydov describes generic singular trajectories surrounding the curves of solutions with two regular horizons~\cite{Davydov:2017zbs}.
Figure~\ref{sm:fig-tuning}(a) shows two examples.
The following proper time asymptotics make their Kasner interpretation explicit.
In an outgoing interior regime with both exponential forces decaying, write
\begin{equation}
 U=-v\tau+O(1)\,,\qquad \phi=w\tau+O(1)\,,\qquad
 v^2+w^2=\Delta^2\,,\qquad v>0\,,\qquad
 -2v+\alpha_a w<0\,.
 \label{sm:kasner-exit}
\end{equation}
Since $-\mathcal R\propto e^{-2\Delta\tau}$, the proper time $t$ to the endpoint obeys $t\propto e^{-(2\Delta-v)\tau}$.
The local metric and scalar have Kasner powers
\begin{equation}
 \begin{gathered}
 \dd s^2\sim-\dd t^2+t^{2p_\parallel}\dd\xi^2
                   +t^{2p_\Omega}\dd\boldsymbol y^2\,,\qquad
 \phi=k_\phi\log(t/t_0)+O(1)\,,\\
 p_\parallel=\frac{v}{2\Delta-v}\,,\quad
 p_\Omega=\frac{\Delta-v}{2\Delta-v}\,,\quad
 k_\phi=-\frac{w}{2\Delta-v}\,,\\
 p_\parallel+2p_\Omega=1\,,\qquad
 p_\parallel^2+2p_\Omega^2+2k_\phi^2=1\,.
 \end{gathered}
 \label{sm:kasner-powers}
\end{equation}
Here, $\boldsymbol y$ denotes local angular coordinates; the factor two in the scalar constraint follows from the normalization $-2(\nabla\phi)^2$.
Nonzero $w$ gives a collapsing singularity in the scalar curvature.
The tuned exit $w=0$, $v=\Delta$ instead has the regular inner horizon behaviour $\text{Rindler}\times S^2$.
The symmetric decay condition $\alpha|w|<2v$ restricts the Kasner exponents, not the coupling to a window.
Thus flowing scalars need not be restricted to the coupling windows admitting two regular horizons.
Reissner--Nordstr\"om exists at every coupling, and the known Toda theories admit regular solutions over a continuous range of the horizon scalar, so singular behaviour at every coupling is not asserted.

\noindent\textit{Relation to Kasner interior mechanics.}\quad
Xiong and L\"u~\cite{Xiong:2026zkb} evaluate the same closed Iyer--Wald surface form between an event horizon and a regulated Kasner singularity, rather than a regular inner horizon.
Their canonical normalization is related to ours by $\phi_\mathrm{XL}=2\phi$ and $F_\mathrm{XL}^a=2F^a$; extending their Maxwell sector to two independent fields gives $Z_a(\phi_\mathrm{XL})=e^{-\alpha_a\phi_\mathrm{XL}/2}$.
For~\eqref{sm:kasner-exit}, their areal radius parameters are
\begin{equation}
 \begin{gathered}
 \Xi=\Delta-v\,,\qquad
 c_1=-\frac{2w}{\Delta-v}\,,\qquad
 l=\frac{2\Delta}{\Delta-v}=1+\frac{c_1^2}{4}\,,\\
 Q_\phi^\mathrm{K}=-\frac w2\,,\qquad
 \Sigma_\mathrm{K}=-\frac{w}{2(\Delta-v)}\,.
 \end{gathered}
 \label{sm:kasner-xl-dictionary}
\end{equation}
These expressions use the same Killing normalization; $\Xi$ and $Q_\phi^\mathrm{K}$ scale with that normalization.
The limit $w\to0$ is nonuniform: $\Xi\to0$, $\Sigma_\mathrm{K}$ diverges, and $p_\Omega\to0$, while the regular inner horizon has finite areal radius.
Consequently their normalized law at the singular endpoint cannot simply be evaluated at $w=0$ to recover our law for two horizons.
The common surface identity applies to both, but the second boundary and its response data differ; a Kasner potential is not an inner horizon entropy.

At fixed couplings, the connection can also be expressed without normalizing by $\Xi$.
In the areal coordinate $\varrho=b$, use $\dd s^2=-h\dd t^2+\dd\varrho^2/f+\varrho^2\dd\Omega_2^2$ and write
$f\sim-F_0\varrho^{-l}$, $h\sim-H_0\varrho^{l-2}$, and
$\phi=k\log\varrho+\phi_\mathrm{K}$, where $l=1+k^2$ and $\Xi=\sqrt{F_0H_0}$.
Evaluating the covariant surface form underlying~\eqref{sm:K} at fixed areal radius cancels the possible logarithm,
$\Xi[k\delta k-\delta l/2]\log\varrho=0$.
The finite surface identity is
\begin{equation}
 \Xi\left[\delta\log\sqrt{F_0}+k\,\delta\phi_\mathrm{K}\right]
 =T_+\delta S_++\sum_a(\Phi_+^a-\Phi_\mathrm{K}^a)\delta Q_a\,,
 \qquad \Phi_\mathrm{K}^a=-A_t^a(0)\,.
 \label{sm:kasner-raw-firstlaw}
\end{equation}
The gauge potentials have finite limits because of the strict inequalities in~\eqref{sm:kasner-exit}.
This is the counterpart of the horizon comparison for two fields and a singular endpoint, with potentials in our sign convention.
It requires a Kasner asymptotic expansion and does not assign an entropy to the singularity.

For comparison, the constant scalar Reissner--Nordstr\"om family has
\begin{equation}
 \dd s_0^2=-f(R)\dd\widehat t^2+
 \frac{\dd R^2}{f(R)}+R^2\dd\Omega_2^2\,,\qquad
 f(R)=1-\frac{2M}{R}+\frac{\Veff^0}{R^2}\,,\qquad \phi=\phi_0\,.
 \label{sm:constant-scalar}
\end{equation}
Its horizons merge at $M=\sqrt{\Veff^0}$ and have $R_+R_-=\Veff^0$.
For arbitrary nonzero symmetric charges, $\Veff=2|Q_1Q_2|\cosh[\alpha(\phi-\phi_0)]$ and $\phi_0=\alpha^{-1}\log|Q_2/Q_1|$.
Shifting the scalar and scaling radii by $\sqrt{|Q_1Q_2|}$ multiplies every entropy by $|Q_1Q_2|$ and leaves $\eta$ unchanged.

\noindent\textit{Geometric obstruction to horizon coalescence.}\quad
Consider a nonconstant translation orbit with two regular horizons.
A smooth merger into a single extremal horizon would require $S_+=S_-$.
The outer constraint and focusing imply
\begin{equation}
 \Veff(\phi_+)=b_+^2-2 b_+b'_+\Delta<b_+^2\,.
 \label{eq:outer}
\end{equation}
Indeed, asymptotic flatness and $b''\le0$ give $b'_+>0$.
Let $k=\Veff(\phi_+)/\Veff^0\ge1$ be the ratio of the outer horizon potential to its attractor value.
Since $S_+-C/S_+$ increases with $S_+$,
\begin{equation}
 S_+-S_->\left(k-\frac{\eta}{k}\right)\Sext>0\,.
 \label{eq:gap}
\end{equation}
Here, $\mathcal F_\phi$ is the focusing integral defined in the Letter, strictly positive for the nonconstant profile.
The last inequality follows from the unsquared focusing identity $b_+b_-=\Veff(\phi_+)-\mathcal F_\phi>0$, which gives $\sqrt\eta<k$.
Focusing thus places the outer entropy threshold $k\Sext$ above the possible coalescence entropy $\sqrt\eta\,\Sext$, even when $\eta>1$.
Both $k$ and $\eta$ are fixed on a translation orbit, so this is a uniform positive bound even for an excess orbit.
The same constraint also bounds the difference of the areal radii,
\begin{equation}
 b_+-b_->\sqrt{\Veff^0}
 \left(\sqrt{k}-\sqrt{\frac{\eta}{k}}\right)>0\,,
 \label{eq:radius-gap}
\end{equation}
while supplying a geometric obstruction to horizon coalescence.
For any deficit family, the weaker bound $S_+-S_->(1-\eta)\Sext>0$ uses only $k\ge1$.

\noindent\textit{The endpoint of a fixed translation orbit.}\quad
We now determine the lower endpoint analytically, keeping the charges, couplings, and nonconstant horizon scalar fixed.
The result concerns the translation orbit of an interior with two regular horizons, rather than every solution at the same charges and couplings.
In particular, it also applies to a nonconstant orbit at fixed horizon scalar with $\eta=1$; such an orbit may lie inside a larger family whose horizon scalar varies.
Regularity and uniqueness of the outer horizon series imply that $\phi_+=\phi_0$ gives the constant scalar solution, so a nonconstant orbit has $\phi_+\ne\phi_0$.

Choose an auxiliary representative with $b_+^2=\Veff(\phi_+)$.
The outer constraint then gives $b'_+=0$.
This representative need only be regular just outside the outer horizon, as constructed in the End Matter, not have an asymptotically flat exterior.
For $\phi_+>\phi_0$, the flux equation implies $\phi'>0$ outside the horizon, and $b''=-b(\phi')^2<0$.
Thus $b'$ becomes negative and stays bounded above by a negative constant, forcing the maximal interval with $b>0$ to end at a finite $r_s>r_+$.
To identify that endpoint, use the exterior variables
\begin{equation}
 \begin{gathered}
 \frac{\dd r}{\dd\tau}=\mathcal R\,,\qquad
 U=\tfrac12\log a^2\,,\qquad
 \ddot U=e^{2U}\Veff\,,\qquad
 \ddot\phi=\tfrac12e^{2U}\partial_\phi\Veff\,,\\
 \dot U^2+\dot\phi^2-e^{2U}\Veff=\Delta^2\,.
 \end{gathered}
 \label{sm:exterior-autonomous}
\end{equation}
Dots denote $\tau$ derivatives, with $\dot U\to\Delta$ and $\dot\phi\to0$ at the outer horizon.
If $\alpha>0$ is the coupling of the growing exponential, log convexity gives
\begin{equation}
 0<\frac{\partial_\phi\Veff(\phi_+)}{\Veff(\phi_+)}
 \le\frac{\partial_\phi\Veff(\phi)}{\Veff(\phi)}<\alpha\,,
 \qquad
 \frac{\ddot\phi}{\ddot U}=\frac{\partial_\phi\Veff}{2\Veff}\,.
 \label{sm:runaway-comparison}
\end{equation}
Integrating these inequalities shows that $U$ and $\phi$ can diverge only together on a finite $\tau$ interval.
If $b$ stayed bounded away from zero, then $U$ would be bounded at finite $r_s$, and so would $\phi$; the autonomous equations would also bound their first derivatives, permitting smooth continuation.
Consequently $b\to0$, $U\to+\infty$, and $\phi\to+\infty$ at $r_s$, at a finite autonomous time $\tau_s$.
For $\phi_+<\phi_0$, apply the same argument to $-\phi$.
Below, $\alpha$ always denotes the coupling of the dominant exponential and $\beta$ the other coupling, both with their signs, so that $\alpha\phi\to+\infty$ in either case.

The orbit map sends infinity to the finite seed radius $R_\lambda$ and has normalization
\begin{equation}
 R_\lambda=r_++\frac{2\Delta}{\lambda-1}\,,\qquad
 c_\lambda=\frac{(\lambda-1)b(R_\lambda)}{2\Delta\sqrt\lambda}\,,\qquad
 \lambda_*:=1+\frac{2\Delta}{r_s-r_+}>1\,.
 \label{sm:orbit-cutoff}
\end{equation}
Every $\lambda>\lambda_*$ has $R_\lambda<r_s$, $c_\lambda>0$, and a regular asymptotically flat completion.
Conversely, an asymptotically flat member has outer radius squared $\lambda\Veff(\phi_+)$, which its outer constraint requires to exceed $\Veff(\phi_+)$, hence $\lambda>1$; a cutoff at or beyond $r_s$ cannot give such a completion.
Thus $(\lambda_*,\infty)$ is the full asymptotically flat interval of this fixed orbit.
Its entropy gap is strictly increasing:
\begin{equation}
 \begin{gathered}
 D(\lambda)=\lambda S_{+0}-\lambda^{-1}S_{-0}\,,\qquad
 D'(\lambda)=S_{+0}+\lambda^{-2}S_{-0}>0\,,\\
 D_* =\Sext\left(\lambda_*k-\frac{\eta}{\lambda_*k}\right)>0\,,
 \end{gathered}
 \label{sm:orbit-infimum}
\end{equation}
where $k=\Veff(\phi_+)/\Veff^0$ and $S_{+0}=k\Sext$.
The endpoint is an infimum, not an attained minimum among regular asymptotically flat members.
Since $\lambda_*>1$, it is strictly above the bound $\Sext(k-\eta/k)$.
Here and below, $D=S_+-S_-$, as in the Letter.
The location $r_s$, and therefore the numerical value of $D_*$, generally still requires exterior integration.

At the cutoff $\tau_c=-(\log\lambda)/(2\Delta)$, the physical observables are
\begin{equation}
 \begin{gathered}
 c=e^{-U(\tau_c)}\,,\qquad
 \widehat M=c\dot U(\tau_c)\,,\qquad
 \widehat T_\pm=\frac{\pm c\Delta}{2S_\pm}\,,\\
 \phi_\infty=\phi(\tau_c)\,,\qquad
 \widehat\Sigma=-c\dot\phi(\tau_c)\,.
 \end{gathered}
 \label{sm:orbit-observables}
\end{equation}
Equivalently, radial data at $R_\lambda$ give
\begin{equation}
 \widehat M=\frac{\lambda+1}{2\sqrt\lambda}b(R_\lambda)
       -\frac{2\Delta\sqrt\lambda}{\lambda-1}b'(R_\lambda)\,,
 \qquad
 \widehat\Sigma=-\frac{2\Delta\sqrt\lambda}{\lambda-1}
                 b(R_\lambda)\phi'(R_\lambda)\,.
 \label{sm:cutoff-charges}
\end{equation}
These identities also fix the normalization needed when the seed exterior itself is not asymptotically flat.

\noindent\textit{Entropy coordinate and runaway scaling.}\quad
Let $S_{\pm*}$ be the finite endpoint entropies and define the dimensionless coordinate and functions
\begin{equation}
 x=\log\frac{S_+}{S_{+*}}=\log\frac\lambda{\lambda_*}
   =2\Delta(\tau_s-\tau_c)\,,\qquad
 h(x)=c\Delta\,,\qquad \varphi(x)=\phi_\infty\,.
 \label{sm:entropy-coordinate}
\end{equation}
Then $S_\pm=S_{\pm*}e^{\pm x}$, and $x\downarrow0$ approaches the infimum.
With subscripts denoting $x$ derivatives,~\eqref{sm:exterior-autonomous} becomes the exact system
\begin{equation}
 \begin{gathered}
 h_{xx}+h\varphi_x^2=\frac h4\,,\qquad
 \varphi_{xx}=\frac{\partial_\varphi\Veff}{8h^2}\,,\qquad
 4h_x^2+4h^2\varphi_x^2-h^2=\Veff\,,\\
 \widehat M=2h_x\,,\quad \widehat\Sigma=2h\varphi_x\,.
 \end{gathered}
 \label{sm:entropy-orbit-system}
\end{equation}
In particular, the first law holds exactly as
$\widehat M_x+\widehat\Sigma\varphi_x=h/2
=\widehat T_+(S_+)_x$.
The scalar work term cancels the divergent leading mass derivative at the runaway boundary.

The dominant exponential can be selected without assuming a power law.
The ratio in~\eqref{sm:runaway-comparison} tends to $\alpha/2$, which first gives $\dot\phi/\dot U\to\alpha/2$ and then $\phi/U\to\alpha/2$.
Set
\begin{equation}
 L=4+\alpha^2\,,\qquad X=2U+\alpha\phi\,,\qquad Y=\alpha U-2\phi\,,
 \qquad \rho=\frac{2\alpha(\alpha-\beta)}L>0\,.
 \label{sm:runaway-variables}
\end{equation}
Thus $Y=o(X)$, while the equations give
\begin{equation}
 \begin{aligned}
 \ddot X&=\frac L2 Q_\alpha^2e^X
       +\frac{4+\alpha\beta}{2}Q_\beta^2
           e^{(1-\rho/2)X+2(\alpha-\beta)Y/L}\,,\\
 \ddot Y&=(\alpha-\beta)Q_\beta^2
           e^{(1-\rho/2)X+2(\alpha-\beta)Y/L}\,.
 \end{aligned}
 \label{sm:runaway-XY}
\end{equation}
Here, $Q_\alpha,Q_\beta$ are the charges of the corresponding exponentials.
The second term in the $X$ equation is smaller than the first by $e^{(\beta-\alpha)\phi}\to0$.
Integrating the leading Liouville equation therefore gives $X=-2\log x+X_*+o(1)$.
Since $Y=o(\log(1/x))$, the remaining force is $x^{\rho-2+o(1)}$.
It is twice integrable for every $\rho>0$, proving that $Y$ has a finite limit rather than assuming this in a dominant balance.
Consequently,
\begin{equation}
 \begin{gathered}
 h\sim A x^p\,,\qquad
 \varphi=-q\log x+\varphi_*+o(1)\,,\\
 p=\frac4{4+\alpha^2}\in(0,1)\,,\qquad
 q=\frac{2\alpha}{4+\alpha^2}\,,\qquad
 Q_\alpha^2e^{\alpha\varphi_*}=4pA^2\,,
 \end{gathered}
 \label{sm:runaway-leading}
\end{equation}
where $A>0$ and $\varphi_*$ are matching constants, and $\varphi_*$ is not the attractor value $\phi_0$.
In the seed radial coordinate this is $b\propto(r_s-r)^p$ and $\phi=-q\log[(r_s-r)/\ell]+O(1)$, for any fixed reference length $\ell$.

The endpoint scaling and several normalization independent ratios follow directly:
\begin{equation}
 \begin{gathered}
 \mathscr S:=D-D_*=(S_{+*}+S_{-*})x+O(x^2)\,,\qquad
 \widehat T_\pm\sim\frac{\pm A}{2S_{\pm*}}x^p\,,\qquad
 \widehat M\sim2Ap\,x^{p-1}\,,\qquad
 \widehat\Sigma\sim-2Aq\,x^{p-1}\,,\\
 \frac{\widehat\Sigma}{\widehat M}\longrightarrow-\frac\alpha2\,,
 \qquad
 \frac{\widehat M^2}{\Veff(\phi_\infty)}\longrightarrow p\,,
 \qquad \widehat M\propto\widehat T_+^{-\alpha^2/4}\,,\\
 \frac{\widehat M\mathscr S}{\widehat T_+}
 \longrightarrow4pS_{+*}(S_{+*}+S_{-*})\,.
 \end{gathered}
 \label{sm:runaway-observables}
\end{equation}
Thus the finite entropy infimum is approached at zero normalized temperature, infinite mass, and runaway modulus; it is not a finite mass extremal limit at fixed modulus.

\noindent\textit{The endpoint in dressed charge units.}\quad
The transformation to a fixed vacuum~\eqref{sm:fixed-vacuum-shift} changes neither the divergent mass nor the vanishing temperature in fixed units.
A further length rescaling exposes a different limiting exterior.
Let $\alpha$ be the dominant coupling and $\beta$ the other coupling, both with their signs, and set
\begin{equation}
 \mathcal Q_\text{d}=|Q_\alpha|e^{\alpha\varphi(x)/2}
       \sim2\sqrt p\,A x^{p-1}\,,\qquad
 \frac{\mathcal Q_\beta}{\mathcal Q_\text{d}}=O(x^{\rho/2})\longrightarrow0\,.
 \label{sm:gm-charge-scale}
\end{equation}
Equations~\eqref{sm:runaway-leading} and~\eqref{sm:runaway-observables} imply
\begin{equation}
 \begin{gathered}
 \frac{\widehat M}{\mathcal Q_\text{d}}\longrightarrow\sqrt p\,,\qquad
 \frac{\widehat\Sigma}{\mathcal Q_\text{d}}\longrightarrow-\frac\alpha2\sqrt p\,,\\
 \frac{S_\pm}{\mathcal Q_\text{d}^2}
 \sim\frac{S_{\pm*}}{4pA^2}x^{2(1-p)}\longrightarrow0\,.
 \end{gathered}
 \label{sm:gm-observables}
\end{equation}
The mass relation is therefore an asymptotic ratio, $\widehat M/\mathcal Q_\text{d}\to1/\sqrt{1+\alpha^2/4}$; both dimensionful quantities diverge.
The entropy product survives this rescaling only as a ratio of vanishing quantities.
Since $\Veff^0$ is unchanged by the vacuum shift, it depends on the charges only through the dressed charges,
\begin{equation}
 \Veff^0=C_{\alpha\beta}\,\mathcal Q_\text{d}^{2(1-w)}\mathcal Q_\beta^{2w}\,,\qquad
 w=\frac{\alpha}{\alpha-\beta}\in(0,1)\,,\qquad
 C_{\alpha\beta}=\left(1-\frac\alpha\beta\right)\left(-\frac\beta\alpha\right)^{w}\,.
 \label{sm:gm-sext}
\end{equation}
With~\eqref{sm:gm-charge-scale}, $\Sext/\mathcal Q_\text{d}^2=O(x^{\rho w})=O(x^{2(1-p)})$, the same power as $S_\pm/\mathcal Q_\text{d}^2$ in~\eqref{sm:gm-observables}, as required by $S_+S_-=\eta\Sext^2$ along the orbit.
The limiting single charge Gibbons--Maeda exterior has zero horizon area and no attractor value, so it retains no memory of $\eta$; the product is recorded only by the two horizon interior.
The shifted horizon scalars behave as
$\widetilde\phi_h=\bar\phi+\phi_h-\varphi_*+q\log x+o(1)$, running opposite to the original asymptotic scalar.

The spacetime limit follows from the exterior reconstruction, rather than these charge ratios alone.
For a member whose infinity is at $x=2\Delta(\tau_s-\tau_c)$, let $z=2\Delta(\tau_s-\tau)>x$ label points in its exterior.
In terms of the same seed functions $h$ and $\varphi$, the exact metric is
\begin{equation}
 \begin{aligned}
 \dd s^2={}&-\left[\frac{h(x)}{h(z)}\right]^2\dd\widehat t^2
 +\frac{h(z)^2}{4\sinh^4[(z-x)/2]}\dd z^2
 +\frac{h(z)^2}{\sinh^2[(z-x)/2]}\dd\Omega_2^2\,,\\
 \widetilde\phi(z)={}&\bar\phi+\varphi(z)-\varphi(x)\,.
 \end{aligned}
 \label{sm:gm-exact-exterior}
\end{equation}
Set $y=z/x$, $\bar t=\widehat t/\mathcal Q_\text{d}$, and $\bar r=1/[\sqrt p(y-1)]$.
Using $h(xy)/(Ax^p)\to y^p$ and $\varphi(xy)-\varphi(x)\to-q\log y$ gives
\begin{equation}
 \begin{gathered}
 \frac{\dd s^2}{\mathcal Q_\text{d}^2}\longrightarrow
 -H^{-2p}\dd\bar t^2
 +H^{2p}(\dd\bar r^2+\bar r^2\dd\Omega_2^2)\,,\qquad
 H=1+\frac1{\sqrt p\,\bar r}\,,\\
 \widetilde\phi\longrightarrow\bar\phi-q\log H\,,\qquad
 p=\frac1{1+\alpha^2/4}\,,\quad q=\frac\alpha2p\,.
 \end{gathered}
 \label{sm:gm-exterior-limit}
\end{equation}
The values and first $y$ derivatives of the rescaled seed functions converge at $y=1$ by~\eqref{sm:runaway-leading} and~\eqref{sm:runaway-observables}.
Smooth dependence of the rescaled differential equations then gives smooth convergence on compact exterior annuli $0<\bar r_1\le\bar r\le\bar r_2<\infty$.
Maxwell's equation fixes the surviving field with unit dressed charge; the other dressed charge vanishes.
This is the extremal single charge Gibbons--Maeda exterior with dilaton coupling $\alpha/2$~\cite{Gibbons:1987ps}.
The convergence does not include its singular surface of zero area at $\bar r=0$.
In particular, the two regular horizons of every finite member do not converge to two regular horizons of this limiting spacetime.

The temperature in units of the growing mass or dressed charge need not vanish:
\begin{equation}
 \widehat T_+\widehat M\sim\frac{pA^2}{S_{+*}}x^{2p-1}\,,\qquad
 \widehat T_+\mathcal Q_\text{d}\sim\frac{\sqrt p\,A^2}{S_{+*}}x^{2p-1}\,,\qquad
 2p-1=\frac{4-\alpha^2}{4+\alpha^2}\,.
 \label{sm:gm-temperature}
\end{equation}
These quantities diverge for $|\alpha|>2$, approach finite nonzero limits for $|\alpha|=2$, and vanish for $|\alpha|<2$.
Thus the symmetric deficit and asymmetric excess examples have $\widehat T_+\to0$ in fixed charge units but $\widehat T_+\widehat M\to\infty$.
There is no contradiction: the comparison length diverges, and convergence of the exterior away from the singular inner boundary does not control the surface gravity of the approximating horizons.

At the opposite, large area end, the shift to $\bar\phi$ remains finite, so the dressed charges remain bounded while $\widehat M\to\infty$.
The regular outer horizon series in~\eqref{sm:gm-exact-exterior}, now with the cutoff tending to the seed horizon, gives the Schwarzschild exterior in mass units, with constant scalar $\bar\phi$.
In particular,
\begin{equation}
 \frac{\mathcal Q_a}{\widehat M}\to0\,,\qquad
 \widehat T_+\widehat M\to\frac1{8\pi}\,,\qquad
 \frac{S_+}{\widehat M^2}\to4\pi\,,\qquad
 \frac{S_-}{\widehat M^2}\to0\,.
 \label{sm:schwarzschild-end}
\end{equation}
Accordingly, the families at fixed vacuum interpolate between Schwarzschild and extremal single charge dilatonic \emph{scaled exteriors}.
Neither description is an endpoint at finite mass of the original orbit at fixed charges, and the runaway limit remains distinct from the regular two charge attractor.

\noindent\textit{Numerical validation of the orbit scaling.}\quad
We test~\eqref{sm:runaway-observables} on both the symmetric deficit and the asymmetric excess examples.
Independent reflection matching gives $\nu=2.99634741419585$ at $\phi_+=0.2$ for the deficit, and $\phi_+=0.44279366227294$ at couplings $(\sqrt7,-2\sqrt7)$ for the excess.
The respective products are $\eta=0.99985996647422$ and $1.02205318801896$.
The matching uses arithmetic with 55 digits, with Taylor order, step size, and initial horizon cutoff varied independently.
A separate DOP853 evolution across the full interior, without reflection during the evolution, confirms three and two scalar nodes and reproduces these products within $5\times10^{-14}$.

For the exterior test, choose $r_\pm=\pm1$, $b_+^2=\Veff(\phi_+)$, and $b'_+=0$.
We integrate the radial equations in the monotone scalar coordinate, evolving $r$, $\log b$, $W=\log(\mathcal R\phi')$, and $\dd\log b/\dd\phi$.
The endpoint distance is computed by direct numerical quadrature,
\begin{equation}
 x=2\int_{\phi_\infty}^{\infty}\!e^{-W(\psi)}\dd\psi\,,\qquad
 \mathscr S=S_{+*}(e^x-1)-S_{-*}(e^{-x}-1)\,,
 \label{sm:scaling-quadrature}
\end{equation}
where $e^{\pm x}-1$ is evaluated without subtractive cancellation.
We truncate the integral at scalar cutoffs $30$, $40$, and $50$ and accumulate its subintervals backwards.
Neither an analytic tail nor a predicted exponent enters the endpoint or gap reconstruction.
The resulting infima are $D_*/\Sext=8.72352956940$ and $5.70556231944$.

The reference exterior runs use twelfth order horizon series at $r-r_+=10^{-4}$ and DOP853 with relative tolerance $3\times10^{-14}$, absolute tolerance $10^{-15}$, and scalar step cap $0.025$.
The scaled Hamiltonian residual is below $6.3\times10^{-14}$ on the sampled exterior.
Varying the tolerance, step cap, and horizon offset changes the measured quantities by less than $4.3\times10^{-13}$ relative over $10^{-12}\le \mathscr S/\Sext\le10^{-4}$.
Increasing the scalar cutoff from $40$ to $50$ changes $x$ by at most $2\times10^{-36}$.
These are empirical convergence checks.

\begin{figure}[tbp]
 \centering
 \includegraphics[width=.96\textwidth]{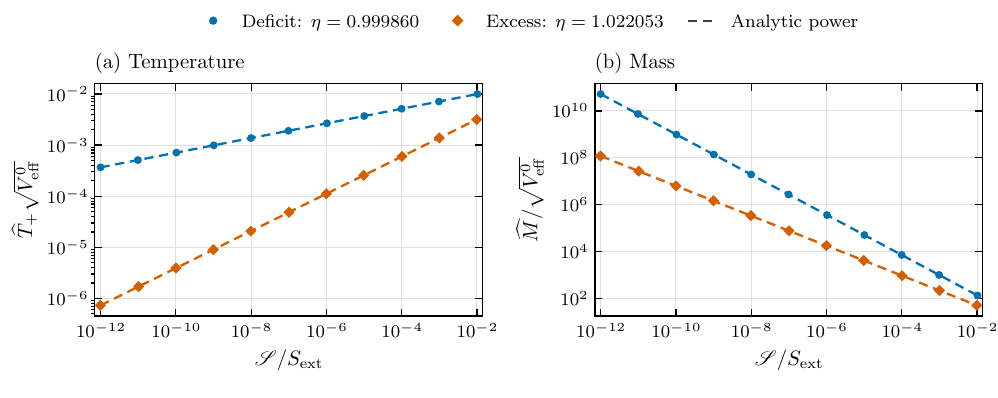}
\caption{\textsf{Numerical validation of fixed coupling orbit endpoint scaling.
Blue circles show the symmetric deficit branch and orange diamonds the asymmetric excess branch, with $\Veff^0=2$ and $3$, respectively.
Here, $\mathscr S=(S_+-S_-)-D_*$, and the endpoint is obtained by direct numerical quadrature.
Markers show a logarithmically spaced subset of the numerical data; dashed curves have the analytic slopes $p$ in (a) and $p-1$ in (b), with only their prefactors fitted on $10^{-12}\le \mathscr S/\Sext\le10^{-8}$.
The predictions are $p=0.143118382558$ for the deficit and $p=4/11$ for the excess.}}
 \label{sm:fig-endpoint-scaling}
\end{figure}

Fitting $(\log[\widehat T_+\sqrt{\Veff^0}],\log[\widehat M/\sqrt{\Veff^0}],\phi_\infty)$ against $\log(\mathscr S/\Sext)$ over $10^{-12}\le \mathscr S/\Sext\le10^{-8}$ gives the respective slope triples
\begin{equation}
 \begin{aligned}
 \text{Deficit}:&\quad(0.143118382518,-0.856881617189,-0.350193533801)\,,\\
 \text{Excess}:&\quad(0.363636363558,-0.636363636225,-0.481045692878)\,.
 \end{aligned}
 \label{sm:scaling-fits}
\end{equation}
The fits use 129 deficit and 177 excess samples; their slopes differ from $(p,p-1,-q)$ by less than $3\times10^{-10}$ and approach these values as the fit window moves toward the endpoint.
At $\mathscr S/\Sext\simeq10^{-12}$, the ratio $\widehat M\mathscr S/[4pS_{+*}(S_{+*}+S_{-*})\widehat T_+]$ differs from unity by less than $6\times10^{-13}$; the limits of $\widehat M^2/\Veff$ and $\widehat\Sigma/\widehat M$ are also reproduced.
Figure~\ref{sm:fig-endpoint-scaling} shows the temperature and mass tests.
The source package includes the scripts, samples, and refinement records.

\begin{figure}[tbp]
\centering
\includegraphics[width=.95\textwidth]{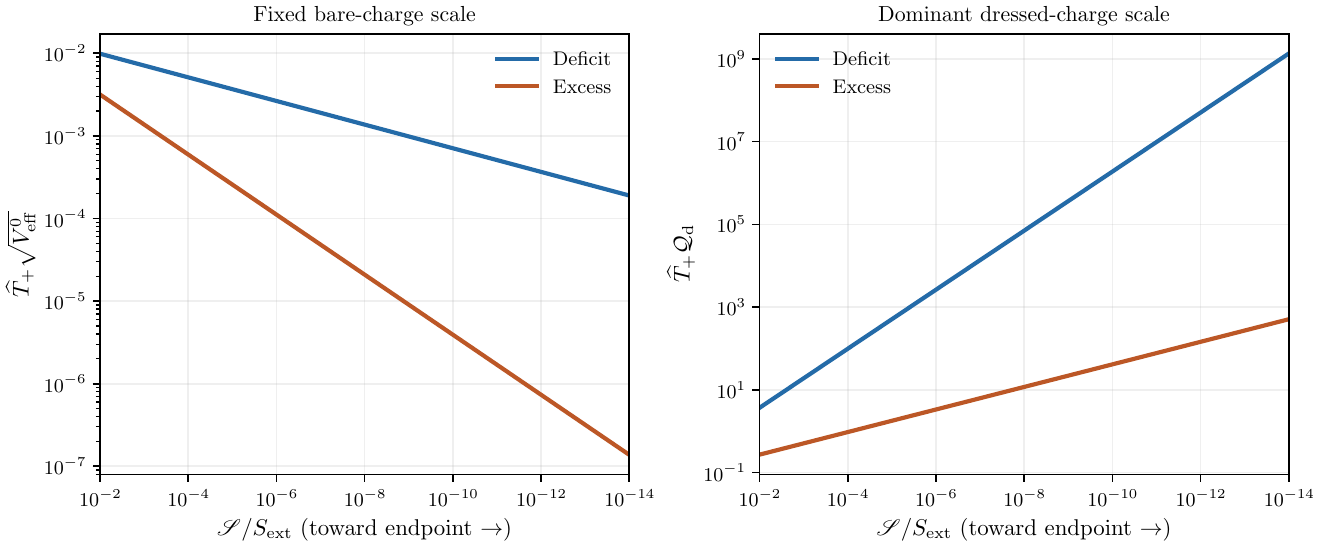}
\caption{\textsf{Temperature in two scales, from the saved numerical exterior
solutions. Approaching the endpoint to the right,
$\widehat T_+\sqrt{\Veff^0}$ vanishes, while
$\widehat T_+\mathcal Q_\text{d}$ diverges for both worked examples,
which have $|\alpha_\text{d}|>2$. The latter has power $2p-1$ in
$\mathscr S/\Sext$.}}
\label{sm:fig-temperature-scales}
\end{figure}

\noindent\textit{Numerical check in the dominant charge scale.}\quad
We postprocess the same exterior solutions used for the endpoint validation;
no additional differential equations are integrated.
Set $\mathcal Q_\text{d}=|Q_\text{d}|e^{\alpha_\text{d}\phi_\infty/2}$ and
$p=4/(4+\alpha_\text{d}^2)$.
The ratios $\widehat M/\mathcal Q_\text{d}$ and
$\widehat\Sigma/\widehat M$ approach $\sqrt p$ and
$-\alpha_\text{d}/2$, respectively.
Table~\ref{sm:table-dressed-validation} gives the maximum relative deviations
over $10^{-12}\le \mathscr S/\Sext\le10^{-8}$, containing 129 deficit and 177 excess
samples.
The fitted powers of $\widehat T_+\widehat M$ are
$-0.713763234671$ and $-0.272727272667$, compared with the respective
predictions $2p-1=-0.713763234884$ and $-0.272727272727$.
The fitted powers of $4S_\pm/\mathcal Q_\text{d}^2$ agree with
$2(1-p)$ within $2.1\times10^{-10}$.
Figure~\ref{sm:fig-temperature-scales} compares the two temperature scales.
The inner entropy is inherited from the matched interior through the exact
product relation.
These records test observable ratios, not independently the convergence of
the full geometry.
The accompanying \texttt{validation/gm\_endpoint/} directory contains the
postprocessing script, derived samples, and audit records.

\begin{table}[tbp]
\centering
\begin{tabular}{lrr}
\toprule
Branch & $\max|\widehat M/(\mathcal Q_\text{d}\sqrt p)-1|$
 & $\max|-2\widehat\Sigma/(\alpha_\text{d}\widehat M)-1|$\\
\midrule
Deficit & $3.46\times10^{-9}$ & $4.04\times10^{-9}$\\
Excess & $1.43\times10^{-9}$ & $2.25\times10^{-9}$\\
\bottomrule
\end{tabular}
\caption{\textsf{Relative deviations from the dominant charge limits in the
stated endpoint window. These measure finite distance from the limit;
the underlying integration refinements are described above.}}
\label{sm:table-dressed-validation}
\end{table}

\noindent\textit{Controlled corrections in the regime of the Letter.}\quad
The branches displayed in the Letter have $\nu>1$, equivalently $\alpha\beta<-4$, and hence $\rho>2$ for either runaway direction.
In this regime the force in the $Y$ equation gives $Y=Y_*+Y_1x+O(x^\rho)$, and the derivative of the Liouville energy of $X$ is $O(x^{\rho-3})$, which is integrable.
Matching these two constants gives an exact resummation of the dominant exponential sector.
Define
\begin{equation}
 \begin{gathered}
 K^2=\frac L{16}-\frac{\chi^2L^2}{4\alpha^2}\,,\qquad
 F_K(x)=\frac{\sinh(Kx)}K\,,\\
 h_0=Ae^{\chi x}F_K(x)^p\,,\qquad
 \varphi_0=\varphi_*+\frac{2\chi}{\alpha}x-q\log F_K(x)\,.
 \end{gathered}
 \label{sm:runaway-envelope}
\end{equation}
Here, $\chi$ is a matching constant, $F_0=x$, and for $K^2<0$ the real continuation is $F_K=\sin(\sqrt{-K^2}x)/\sqrt{-K^2}$.
These formulas are local near $x=0$, where $F_K>0$.
They solve~\eqref{sm:entropy-orbit-system} exactly after retaining only the dominant exponential: in particular,
$h_{0,xx}/h_0+(\varphi_{0,x})^2=\chi^2L/\alpha^2+pK^2=1/4$.
The first correction from the other exponential is
\begin{equation}
 \begin{aligned}
 \frac h{h_0}&=1-\frac{H}{\rho^2-1}x^\rho+O(x^{\rho+1})\,,\\
 \varphi-\varphi_0&=
 \frac{H(2\alpha\beta-L)}{4\alpha(\rho^2-1)}x^\rho
       +O(x^{\rho+1})\,,\qquad
 H=\frac{Q_\beta^2e^{\beta\varphi_*}}{4A^2}\,.
 \end{aligned}
 \label{sm:runaway-correction}
\end{equation}
Linearizing~\eqref{sm:entropy-orbit-system} about the envelope fixes these coefficients; the next forcing has order $x^{\rho-1}$ in the second order equations.
There is no resonance for integer $\rho>2$.
For example, the resulting quadratic expansion is
\begin{equation}
 \begin{aligned}
 h&=Ax^p\left[1+\chi x+
       \left(\frac1{24}+\frac{\alpha^2-2}{3\alpha^2}\chi^2\right)x^2
       +O\!\left(x^{\min(3,\rho)}\right)\right]\,,\\
 \varphi&=\varphi_*-q\log x+\frac{2\chi}{\alpha}x+
       \left(\frac{L\chi^2}{12\alpha}-\frac\alpha{48}\right)x^2
       +O\!\left(x^{\min(3,\rho)}\right)\,.
 \end{aligned}
 \label{sm:runaway-quadratic}
\end{equation}
Outside this regime, the leading result~\eqref{sm:runaway-leading} still holds: its first relative correction in $h$, and additive correction in $\varphi$, is $O(x^\rho)$ for $0<\rho<1$, $O(x|\log x|)$ at $\rho=1$, and $O(x)$ for $\rho>1$.
At $\rho=2$, the apparent resonance cancels because $4+\alpha\beta=0$ in~\eqref{sm:runaway-XY}; no $x^2\log x$ term is required.
These statements concern the first correction, not the absence of higher order logarithms at other couplings.

\noindent\textit{The limiting geometry and asymptotic normalization.}\quad
At $\lambda=\lambda_*$, the orbit map still gives a smooth metric on and between both horizons and at every finite exterior radius, if its time coordinate is kept unnormalized.
The seed singularity is moved to infinity, where
$b_{\lambda_*}\propto r^{1-p}$, $a_{\lambda_*}^2\propto r^{2p}$, and $\phi_{\lambda_*}=q\log(r/\ell)+O(1)$.
In terms of the areal radius $\varrho=b_{\lambda_*}$, a constant time rescaling gives
\begin{equation}
 \dd s_*^2\sim
 -\left(\frac{\varrho}{\varrho_0}\right)^{8/\alpha^2}\dd T^2
 +\frac{\dd\varrho^2}{(1-p)^2}+\varrho^2\dd\Omega_2^2\,,
 \qquad
 \phi_*\sim\frac2\alpha\log\frac{\varrho}{\varrho_0}+\phi_c\,.
 \label{sm:critical-geometry}
\end{equation}
The constants obey $Q_\alpha^2e^{\alpha\phi_c}=p\varrho_0^2$.
This is not asymptotically flat.
Its proper radial distance to infinity diverges, while the Ricci scalar is
$R\sim8(1-p)^2/(\alpha^2\varrho^2)$ and the Kretschmann scalar is $O(\varrho^{-4})$.
Thus a curvature singularity of this unnormalized limiting exterior is not required.
What fails along the asymptotically flat sequence is its normalization: $c=h/\Delta\to0$, while the asymptotic scalar and ADM mass diverge.
The normalized coordinate separation of the horizons vanishes, but their unequal limiting areas and the divergent field derivatives prevent a smooth extremal merger.
For example, regularity at the outer horizon gives
$\partial_{\widehat r}\phi|_+=\partial_\phi\Veff(\phi_+)/(4hb_+^2)$.
A radial infaller of fixed energy $E$ relative to normalized infinity therefore has a scalar contribution
$2E^2[\partial_\phi\Veff(\phi_+)/(4hb_+^2)]^2$
to $R_{\mu\nu}u^\mu u^\nu$ at that horizon, which diverges as $h^{-2}$.
This is a divergence for that physically fixed family of observers; it does not turn the regular horizon of the differently normalized critical metric into an intrinsic curvature singularity.

\noindent\textit{Numerical continuation to zero temperature.}\quad
For the symmetric example~\eqref{sm:deficit-point}, numerical exterior continuation gives an asymptotically flat interval $b_+>b_+^\text{crit}$ with
\begin{equation}
 b_+^\text{crit}\simeq4.203969176\,,\qquad
 \inf_\text{orbit}\frac{S_+-S_-}{\Sext}\simeq8.72352957\,.
 \label{sm:orbit-endpoint}
\end{equation}
The gap increases with $b_+$ because $S_+=\pi b_+^2$ and $S_-=C/S_+$.
The lower endpoint is the scalar runaway described above.
For a reference radius $b_+^\text{ref}$ and a corresponding choice of autonomous time origin, the orbit samples the exterior trajectory at $s=-\log(b_+/b_+^\text{ref})/\Delta$, giving $b_+^\text{crit}=b_+^\text{ref}e^{-s_\text{crit}\Delta}$.
The physical quantities are obtained from~\eqref{sm:orbit-observables}.
As $b_+$ decreases to $b_+^\text{crit}$, the modulus, normalized mass, and scalar charge diverge, while the outer temperature tends to zero.
As $b_+\to\infty$, $\widehat M\sim b_+/2$, $\widehat T_+\sim1/(4\pi b_+)$, $\phi_\infty\to0.2$, and $S_-\to0$ while $S_+\to\infty$.
Neither endpoint is a finite mass extremal limit.

\begin{figure}[tbp]
 \centering
 \includegraphics[width=.96\textwidth]{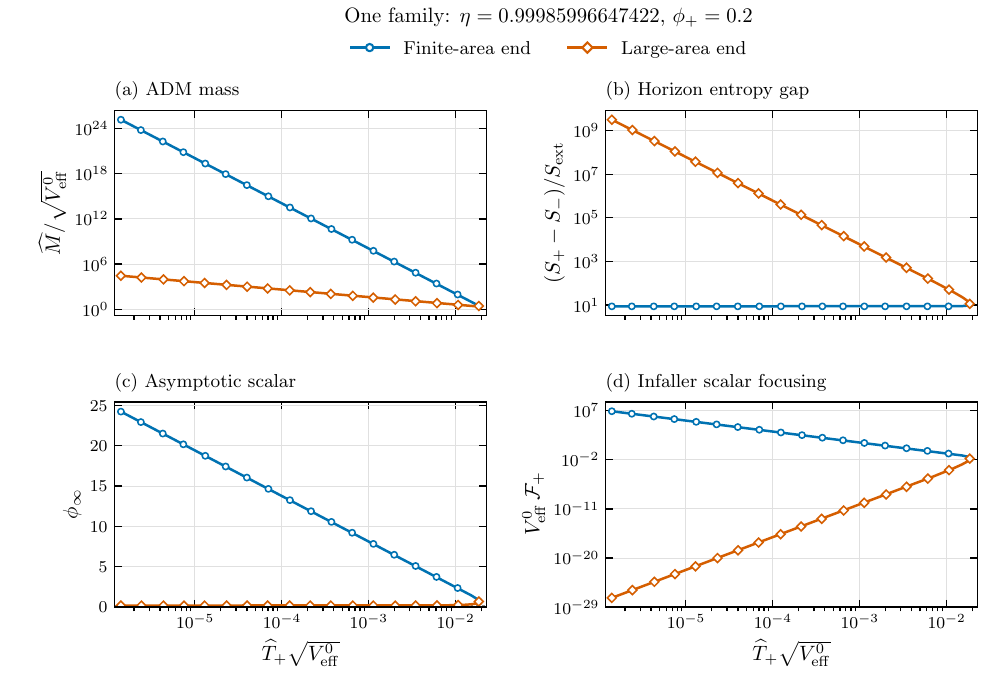}
\caption{\textsf{Numerical continuation to both zero temperature ends of the same symmetric deficit family, with $\Veff^0=2$.
Blue circles approach the finite area runaway boundary; orange diamonds approach the large area end.
The curves join at the temperature maximum and represent different values of the boundary modulus, not two phases at fixed modulus.
Panels show (a) mass, (b) the total horizon entropy gap, (c) the asymptotic scalar, and (d) the scalar contribution $\mathcal F_+$ to the Ricci curvature measured by a radial infaller with energy $E=1$ at infinity.
Lines connect numerical samples; markers show a subset.
The scalar contribution in (d) is evaluated from~\eqref{sm:cold-curvature}.}}
 \label{sm:fig-cold-family}
\end{figure}

Figure~\ref{sm:fig-cold-family} follows both cold ends of this one family, with $Q_1^2=Q_2^2=1$, $\phi_+=0.2$, and $\eta=0.99985996647422$, down to $\widehat T_+=10^{-6}$.
The 681 samples use direct exterior integration and, for cutoffs closer to the seed horizon than the integration start, the regular horizon series.
No extrapolation of the exterior interpolant is used.
At eight prescribed temperatures we separately solve for the two orbit members.
Changing the integration tolerance, step cap, horizon offset, and series order changes their observables by less than $3.3\times10^{-12}$ relative; the scaled asymptotic constraint residual is below $2.7\times10^{-14}$ over the sweep.
The product is inherited from the independently matched interior by the exact orbit map; its preservation in this sweep is not a new interior regularity test.

At the finite area end, fits over $10^{-6}\le\widehat T_+\le10^{-4}$ give
\begin{equation}
 \frac{\dd\log\widehat M}{\dd\log\widehat T_+}\simeq-5.9872226204\,,\qquad
 \frac{\dd\phi_\infty}{\dd\log\widehat T_+}\simeq-2.4468801810\,,
 \label{sm:cold-fits}
\end{equation}
consistent with $-\alpha^2/4$ and $-\alpha/2$ at $\alpha=\sqrt{2\nu(\nu+1)}$.
Logs here are taken in the fixed charge units of this example; using the dimensionless axes of Figure~\ref{sm:fig-cold-family} gives identical slopes.
The limiting entropies are $S_{+*}\simeq55.52248798$ and $S_{-*}\simeq0.71093517$.
For example, at $\widehat T_+=10^{-6}$ we obtain $\widehat M\simeq2.09193\times10^{25}$ and $\phi_\infty\simeq24.22463$.
At this temperature the other member has $\widehat M\simeq3.97887\times10^4$, $S_+\simeq1.98944\times10^{10}$, $S_-\simeq1.98412\times10^{-9}$, and $\phi_\infty\simeq0.200000000883$.
On that end the fitted mass power is $-1.0000000000$, as predicted by $\widehat M\sim(8\pi\widehat T_+)^{-1}$.
The finite area gap becomes indistinguishable from its infimum in ordinary floating point subtraction; resolving $\mathscr S=D-D_*$ there requires~\eqref{sm:scaling-quadrature}.

Horizon curvature and the response of infalling observers distinguish the two limits further.
At each regular member, the horizon constraint and $(a^2b^2)''=2$ give
\begin{equation}
 K_h=\frac{12}{b_h^4}-\frac{24\Veff(\phi_h)}{b_h^6}
             +\frac{20\Veff(\phi_h)^2}{b_h^8}\,,\qquad
 \mathcal F_+:=\left.2(u^\mu\partial_\mu\phi)^2\right|_+
   =\frac{[\partial_\phi\Veff(\phi_+)]^2}{8h^2b_+^4}\,,
 \label{sm:cold-curvature}
\end{equation}
where $h=c\Delta$, $K_h$ is the Kretschmann scalar, and $u^\mu$ is radial timelike infall with energy $E=1$ relative to normalized infinity.
Here, $\mathcal F_+$ is the scalar contribution to $R_{\mu\nu}u^\mu u^\nu$; the Maxwell contribution is $\Veff(\phi_+)/b_+^4$.
Evaluating these identities on the numerical horizon data gives $K_+\to0.0271060$, $K_-\to6.42805\times10^4$, and $\mathcal F_+\sim4.05944\times10^{-6}\widehat T_+^{-2}$ at the finite area end.
The divergence concerns the observers at fixed energy, as discussed above, while the unnormalized critical metric has regular horizons.
At the large area end, $K_+\propto\widehat T_+^4\to0$ but $K_-\propto\widehat T_+^{-8}\to\infty$: the inner horizon collapses.
These curvature values use the exact horizon identities, rather than an independent numerical differentiation of the metric.

\noindent\textit{Orbit extrema.}\quad
The temperature maximum is $\widehat T_+\simeq0.01321792$ at $b_+\simeq4.783371$, whereas the mass minimum is $\widehat M\simeq3.409746$ at $b_+\simeq5.460296$.
The corresponding asymptotic moduli are $0.701020$ and $0.499778$, respectively.
Since the scalar boundary data vary along the orbit, these extrema do not establish a thermodynamic phase transition or a heat capacity singularity in an ensemble with fixed modulus.
The scalar work term in the first law is essential when following this orbit.
Their uniqueness also follows from~\eqref{sm:entropy-orbit-system}.
The magnitude $|\varphi_x|$ decreases strictly from infinity to zero, so $\widehat M_x=2h(1/4-\varphi_x^2)$ changes sign exactly once.
For the temperature, $z=h_x/h$ decreases through any crossing of $z=1$, since $z_x=1/4-\varphi_x^2-z^2<0$ there.
Its endpoint values are $z\sim p/x\to\infty$ and $z\to1/2$, giving exactly one temperature maximum.

\begin{figure}[tbp]
 \centering
 \includegraphics[width=.78\textwidth]{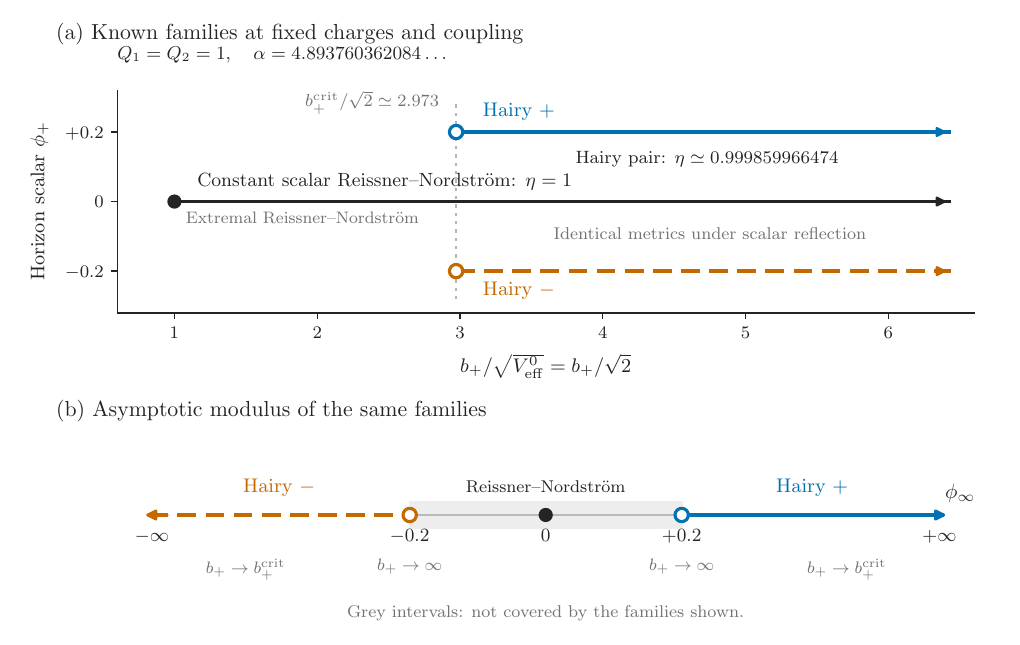}
\caption{\textsf{Known families at the coupling of~\eqref{sm:deficit-point}.
The Reissner--Nordstr\"om family has $\eta=1$; the two families related by scalar reflection have the same deficit and $\phi_+=\pm0.2$.
The nonconstant families end at runaway boundaries rather than a finite mass extremal solution.
Their projected domains in asymptotic moduli space do not classify all hot solutions.
Extremal exteriors exist at every finite modulus in this symmetric model.}}
 \label{sm:fig-families}
\end{figure}

\noindent\textit{Quantitative gap and limiting entropy.}\quad
For the symmetric $\epsilon=0.2$ example, the bound~\eqref{eq:gap} gives $(S_+-S_-)/\Sext>0.859985$, while continuation through the full exterior gives the stronger lower bound in~\eqref{sm:orbit-endpoint}.
At small $\epsilon$, the analytic lower bound starts at $24\epsilon^2\Sext$, whereas $1-\eta$ starts at fourth order.
Writing $D=S_+-S_->0$ gives $S_\pm=(\sqrt{D^2+4C}\pm D)/2$.
If $D\to0$ is accessible, the common entropy is necessarily $\Sinf=\sqrt C$.
This algebraic identification neither fixes $\Sinf$ to $\Sext$ nor guarantees convergence to a smooth extremal spacetime.
The positive lower bound excludes that limit on the orbits considered; a larger family whose horizon scalar varies need not have a uniform positive lower bound.

\noindent\textit{Extremal exteriors at arbitrary finite modulus.}\quad
The separation of the hot families from the extremal solution does not remove points from the extremal attractor basin.
For the symmetric model, a first order construction~\cite{Ceresole:2007wx,Andrianopoli:2007gt} uses the even function
\begin{equation}
 \mathcal W^2+\mathcal W_\phi^2=2\cosh(\alpha\phi)\,,\qquad
 \mathcal W=\sqrt2[1+\nu\phi^2/2+O(\phi^4)]\,.
 \label{sm:superpotential}
\end{equation}
To obtain its local analytic branch, write $\mathcal W=\sqrt2(1+\phi^2y)$.
Then $\phi y'=\sqrt{F-2y-\phi^2y^2}-2y$, where $F=[\cosh(\alpha\phi)-1]/\phi^2$ and $y(0)=\nu/2$.
The derivative of the right hand side with respect to $y$ is $-2-1/\nu$ at the origin, fixing the regular branch.
For $\phi>0$, choose $\mathcal W_\phi=\sqrt{\Veff-\mathcal W^2}>0$.
The radicand cannot first vanish at positive $\phi$: its derivative there would be $\Veff'>0$.
Both $\mathcal W$ and its derivative remain bounded on compact scalar intervals, so the solution extends to every finite $\phi$; reflection covers negative values.

With isotropic radius $\bar r$, $x=1/\bar r$, $H(0)=1$, and $\phi(0)=\phi_\infty$, the flow is
\begin{equation}
 H_x=\mathcal W\,,\qquad \phi_x=-\mathcal W_\phi/H\,,\qquad
 \dd s^2=-H^{-2}\dd t^2+H^2(\dd\bar r^2+\bar r^2\dd\Omega_2^2)\,.
 \label{sm:extremal-flow}
\end{equation}
For positive $\phi_\infty$, $\phi$ decreases and $1+\sqrt2x\le H\le1+\mathcal W(\phi_\infty)x$.
If the scalar approached a positive limit, its integrated decrease would be unbounded; hence $\phi\to0$ and $H/x\to\sqrt2$.
The entropy is $\Sext=2\pi$, with $M=\mathcal W(\phi_\infty)$ and $\Sigma=-\mathcal W_\phi(\phi_\infty)$.
Thus $M^2+\Sigma^2=\Veff(\phi_\infty)$.
This branch approaches constant scalar Reissner--Nordstr\"om as $\phi_\infty\to0$.

The horizon behaviour $\phi\sim A\bar r^\nu$ requires a regularity qualification~\cite{Cremonini:2023vwf}.
At the illustrative $\nu\simeq2.9963$, the scalar is $C^2$ but generically not $C^3$, while curvature and the classical equations are well defined.
For $0<\nu<1$, $R_{\bar r\bar r}=2(\phi')^2$ diverges in ingoing coordinates.
The comparison therefore does not imply that extremal horizons with a running scalar are infinitely smooth at all couplings.
The nondegenerate horizons constructed in the Letter are analytic.

\noindent\textit{Scope and dynamics.}\quad
The area gaps characterize stationary solutions and coexist with the zero temperature limits at divergent mass or modulus described above.
Cauchy horizons can be destabilized by mass inflation~\cite{Poisson:1990eh}; dynamical stability of these families remains to be determined.
The coupling sign rule~\eqref{sm:scaling} holds on regular segments parametrized by $s$, including both sides of the fold in Section~\ref{sm:endpoints}.
An embedding of the couplings of the $\eta\ne1$ families into ungauged $\mathcal N=2$ supergravity also remains to be established.

\section{A related near-horizon entropy comparison}
\label{sm:near-horizon}

In the frame $f_1=e^{-\alpha_1\phi}$, $f_2=e^{\alpha_2\phi}$, set $\mathcal F=-a^2>0$ between the horizons.
Retaining the real profiles $b(r)$ and $\phi(r)$ gives
\begin{equation}
 \begin{gathered}
 \dd s_\text{NH}^2=b^2\,\dd s_{\text{AdS}_2}^2
          +\frac{\dd r^2}{\mathcal F}+\mathcal F\dd\varphi^2\,,\\
 F^1_\text{NH}=\frac{Qe^{\alpha_1\phi}}{b^2}\dd r\wedge\dd\varphi\,,
 \qquad F^2_\text{NH}=P\,\text{vol}_{\text{AdS}_2}\,.
 \end{gathered}
 \label{sm:continuation}
\end{equation}
Here, $\dd s^2_{\text{AdS}_2}$ and $\text{vol}_{\text{AdS}_2}$ are the unit radius AdS$_2$ metric and volume form, and $\varphi$ is an azimuthal angle whose period is fixed below.
For the orientation in which $\text{vol}_{\text{AdS}_2}\wedge\dd r\wedge\dd\varphi$ is positive, the dual fields are $G_1=Q\,\text{vol}_{\text{AdS}_2}$ and $G_2=-Pe^{\alpha_2\phi}\dd r\wedge\dd\varphi/b^2$.
Writing $\mathscr P=\mathcal F b^2=-\mathcal R$,  with primes denoting $r$ derivatives, the polar equations are $\mathscr P''=-2$, $b''/b=-(\phi')^2$, and $(\mathscr P\phi')'=-\partial_\phi\Veff/(2b^2)$.
The polar constraint is the original radial constraint with $\mathcal R=-\mathscr P$.
The map therefore preserves the equations with real fields and unchanged Maxwell kinetic signs.

Both poles are smooth precisely when their required azimuthal periods agree.
The orbit map with $\lambda=b_-/b_+$ uniquely balances the endpoint radii at $\sqrt{\mathcal A}$, where $\mathcal A=b_+b_-$.
Its period is $2\pi\mathcal A/\Delta$.
For a nonconstant scalar this balanced parent has no asymptotically flat exterior, since concavity and equal endpoint radii imply $b'_+<0$.
Only its compact interhorizon region is used.

The horizon area element is $\dd r\,\dd\varphi$, so $S_\text{NH}=\pi\mathcal A=\sqrt{S_+S_-}$.
Define $w_1=-\alpha_2/(\alpha_1-\alpha_2)$ and $w_2=\alpha_1/(\alpha_1-\alpha_2)$.
The first two identities in~\eqref{sm:averages} give the continued magnetic and electric fluxes
\begin{equation}
 p=\frac{w_1\mathcal A}{Q}\,,\qquad
 q=-\frac{w_2\mathcal A}{P}\,,\qquad
 \min_\phi\left(p^2e^{-\alpha_1\phi}+q^2e^{-\alpha_2\phi}\right)
 =\frac{\mathcal A^2}{\Veff^0}\,.
 \label{sm:continued-flux}
\end{equation}
These fluxes generally differ from the original charges.
At these charges the spherical and warped entropies satisfy
\begin{equation}
 S_\text{sph}=\eta\Sext\,,\qquad
 S_\text{NH}=\sqrt\eta\,\Sext\,,\qquad
 \frac{S_\text{NH}}{S_\text{sph}}=\eta^{-1/2}\,.
 \label{sm:entropy-ratio}
\end{equation}
The warped entropy is larger for $\eta<1$ and smaller for $\eta>1$.
Both geometries are extrema of the axisymmetric entropy functional~\cite{Sen:2005wa,Astefanesei:2006dd}.
This compares two near-horizon extrema, not a hot solution and its extremal endpoint.
A global extremal embedding does not follow from a local near-horizon solution~\cite{Kunduri:2013gce}.

At fixed polar metric and fluxes, solve Maxwell's equations to test the symmetric saddle.
Write the SO(2,1)$\times$U(1) invariant metric~\cite{Astefanesei:2007bf,Kunduri:2013gce} as $v(z)\dd s^2_{\text{AdS}_2}+n(z)^2\dd z^2+w(z)^2\dd\psi^2$ with $\psi$ of period $2\pi$.
The scalar dependent part of the reduced entropy functional is
\begin{equation}
 \mathcal E_\text{red}[\phi]
 =\frac\pi2\int_{-1}^{1}\!\dd z\,\frac{vw}{n}(\phi')^2
 +2\pi\left(\frac{p^2}{I_+}+\frac{q^2}{I_-}\right)
 +\mathcal E_\text{grav}\,,\qquad
 I_\pm=\int_{-1}^{1}\!\dd z\,\frac{nw}{v}e^{\pm\alpha\phi}\,.
 \label{sm:entropy-functional}
\end{equation}
The term $\mathcal E_\text{grav}$ is constant along these paths.
At the round seed, a zero mean Legendre variation gives
\begin{equation}
 \frac{\delta^2\mathcal E[P_\ell]}{\pi\mathcal A}
 =\frac{2}{2\ell+1}\left[\ell(\ell+1)-\frac{\alpha^2}{2}\right]\,,
 \qquad \ell\ge1\,.
 \label{sm:hessian}
\end{equation}
At $\alpha^2=24$, the $\ell=1,2$ directions are negative and every $\ell\ge4$ is positive.
Analytic dependence preserves both signs along the deficit branch at small nonzero $\epsilon$.
At $\epsilon=0.2$, the normalized $P_1$ and $P_4$ curvatures are approximately $-7.061815$ and $1.553137$.
Their signs persist under the profile, quadrature, and finite difference refinements recorded in the code.
This is a saddle of the entropy functional, not a statement about Lorentzian stability.
The separate integrable $\alpha=2$ family realizes the $P_1$ zero mode nonlinearly as a physical flat direction.
In the frame $f_1=f_2=e^{-2\phi}$, choose $Q,P>0$ and set $\Gamma_0=2QP$, $|h|<1$, and $\psi\sim\psi+2\pi$.
Its metric and scalar are
\begin{equation}
 \begin{gathered}
 \Gamma(z)=\Gamma_0\frac{1-h^2z^2}{1-h^2}\,,\qquad
 \dd s^2=\Gamma(z)\dd s_{\text{AdS}_2}^2+
 \frac{\Gamma(z)}{1-z^2}\dd z^2+
 \frac{\Gamma_0^2(1-z^2)}{\Gamma(z)}\dd\psi^2\,,\\
 \phi=\frac12\log\frac{P(1-hz)}{Q(1+hz)}\,,\qquad -1\le z\le1\,.
 \end{gathered}
 \label{sm:flat-family}
\end{equation}
Together with $F^2=P\,\text{vol}_{\text{AdS}_2}$ and $G_1=Q\,\text{vol}_{\text{AdS}_2}$, the horizon fluxes are
\begin{equation}
 F^1\big|_H=\frac{P(1-h^2)}{(1+hz)^2}\dd z\wedge\dd\psi\,,\qquad
 G_2\big|_H=-\frac{Q(1-h^2)}{(1-hz)^2}\dd z\wedge\dd\psi\,.
 \label{sm:flat-flux}
\end{equation}
The substitution $r=\Gamma_0z$, $b^2=\Gamma$, and $\mathscr P=\Gamma_0^2(1-z^2)$ verifies the polar equations and constraint, while $\Gamma(\pm1)=\Gamma_0$ ensures smooth poles.
The fluxes $p=P$, $q=-Q$, and entropy $S_\text{NH}=\pi\Gamma_0=2\pi QP$ are independent of $h$.
Unordered pole scalar values distinguish $|h|$ physically; $h=0$ is round with scalar tangent in the $P_1$ direction.
This local flat direction does not establish nonuniqueness of complete asymptotically flat extremal black holes.

\bibliography{ref}